\documentclass[a4paper,12pt]{article}
\pdfoutput=1
\usepackage{epsfig,graphicx,xcolor,amsbsy,amssymb,latexsym,amsfonts,amsmath,,tcolorbox,setspace}
\usepackage{pstricks}
\usepackage{color}
\usepackage{soul}
\usepackage{placeins}
\usepackage{wrapfig,framed,caption}
\usepackage[makeroom]{cancel}

\usepackage{wasysym}

\usepackage{multirow}

\definecolor{shadecolor}{gray}{0.925}

\usepackage{eurosym}
\usepackage[a4paper]{geometry}
\usepackage{hyperref}
\usepackage{graphicx}
\usepackage{wrapfig}
\usepackage{tikz}
\usetikzlibrary{decorations.pathreplacing,decorations.markings,snakes}
\usetikzlibrary{decorations.pathmorphing}
\usetikzlibrary{patterns}
\usetikzlibrary{calc}
\usepackage{xcolor}
\usepackage{amsmath,amssymb,amsfonts,pstricks,setspace}
\usepackage[enableskew]{youngtab}

\usepackage[style=numeric-comp,sorting=none,maxbibnames=9,backend=bibtex]{biblatex}
\numberwithin{equation}{section}

\usepackage{wrapfig}

\newcommand{\bea}{\begin{eqnarray}\displaystyle}
\newcommand{\eea}{\end{eqnarray}}

\newcommand{\zh}{z_{H}}

\newcommand{\tco}{\mathfrak{t}}
\newcommand{\xco}{x}
\newcommand{\tcop}{\mathfrak{t}'}

\newcommand{\rh}{r_H}

\newcommand{\ri}{r_{\text{isco}}}

\newcommand{\eps}{\mathcal{E}}
\newcommand{\fun}{\mathcal{F}}

\newcommand{\sh}{\mathfrak{q}}

\newcommand{\scal}{\phi}

\newcommand{\scalv}{V_{\text{scal,0}}}

\newcommand{\con}{\mathcal{C}}

\newcommand{\coff}[2]{\mathfrak{u}^{(#1)}_{#2}}

\newcommand{\tpo}{\mathfrak{t}_{\text{ps},0}}

\newenvironment{linesmall}[1]
  {\arraycolsep=3pt\scriptsize% change to whatever you need
   \array{#1}}
  {\endarray}

\title{
{\bf Linear Response to Black Hole Deformations}\\[10pt]}

\author{\large\textsc{Stefan~Hohenegger\footnote{\tt s.hohenegger@ip2i.in2p3.fr}}}

\begin{document}

\maketitle
\thispagestyle{empty}

\maketitle
\thispagestyle{empty}
${}$\\
\begin{center}
\renewcommand{\thefootnote}{\fnsymbol{footnote}}\vspace{-0.5cm}
${}^{\footnotemark[1]}$ Université Lyon 1, CNRS, IP2I, UMR 5822, Villeurbanne, France\\[3.2cm] 
\end{center}

\begin{abstract}
Modifications of gravity beyond (vacuum) General Relativity have a subtle impact on the physics around black holes. Describing these effects across different observables is, however, a priori strongly model dependent. While effective descriptions of the geometry allow to efficiently capture deviations from benchmark predictions, they sometimes lack a clear passage towards (effective) new physics. Inspired by linear response theory, in this work I therefore propose a formalism that allows to calculate leading corrections of observables relative to the Schwarzschild black hole anywhere in space-time (outside of the horizon): the order of corrections is fixed through an eigenvalue $\eps$ of the Einstein tensor, which is identified with an external energy momentum tensor that generates the deformation of the geometry. In this ways, $\eps$ can be linked to effective physics models. Focusing on spherically symmetric and static black holes, I illustrate the approach at the hand of a number of observables in the vicinity of the horizon, such as characteristic quantities of particles moving around the black hole (\emph{e.g.} the black hole shadow or the radius of the innermost stable circular orbit) or the spectrum of scalar quasi-normal modes. Reversing the logic, I argue that the knowledge of such observables provides non-trivial constraints on $\eps$, which in turn grant insights into the effective physics in the vicinity of the black hole.

\end{abstract}

\newpage

\tableofcontents

\section{Introduction}
Reconciling gravity with the other fundamental interactions in nature is a notoriously difficult task. While numerous theoretical approaches have been developed and explored during the last decades, recent (\emph{e.g.} gravitational wave experiments such as LIGO-Virgo-KAGRA \cite{Abbott:2016blz,Abbott:2016nmj,Abbott:2019niy,LIGOScientific:2020tif}, the Event Horizon Telescope \cite{EventHorizonTelescope:2019dse,EventHorizonTelescope:2021srq,EventHorizonTelescope:2021bee,EventHorizonTelescope:2022urf,EventHorizonTelescope:2022wkp,EventHorizonTelescope:2022xqj} or the Planck satellite mission\footnote{For a list of Planck publications see \url{https://www.cosmos.esa.int/web/planck/publications}.}) and future (\emph{e.g.} the Einstein Telescope or future space-borne missions such as LiteBIRD or LISA) experimental efforts produce more and more accurate data. These provide tests of General Relativity (GR) and potentially allow for glimpses at new physics beyond it, potentially of a quantum nature. Central objects of study in such endeavours are \emph{black holes}: while the \emph{Schwarzschild} space-time \cite{Schwarzschild:1916uq} is a well-known classical solution of vacuum GR, many other geometries have been proposed in the context of theories that go beyond the latter. Indeed, a number of asymptotically flat and static black holes that are spherically symmetric and have a non-extremal event horizon have been studied in the literature \cite{Dymnikova:1992ux,Kazakov:1993ha,Garcia95,Dymnikova:2004qg,Hayward:2005gi,Simpson:2018tsi,Simpson:2019mud}.

Given this variety of black hole geometries stemming from theories beyond vacuum GR, it is an important question how to compare these theoretical models to current and future experimental results and how to extract potentially new physics (that is incorporated in these model) from such measurements. In this work, we therefore propose an \emph{effective description} of the space-time outside of a black hole, we explain how to use such a description to express observables in a universal fashion and demonstrate how measuring various different such observables can provide direct insights into (effective) physics beyond GR. We stress that our approach is not a proposal for a fundamental theory that predicts black hole geometries. Instead, we provide a framework to uniformly describe such geometries and render them comparable among themselves as well as with measured observables, which in turn shall allow us to extract model-independent features of gravitational physics beyond the vacuum GR benchmark. 

Concretely, in order to demonstrate the viability of our approach, we focus in this work on a class of static, spherically symmetric and asymptotically flat black holes with a non-extremal event horizon. Model-independent parametrisations of such geometries generalising the Schwarzschild space-time, have previously been studied (see \emph{e.g.} \cite{642b3682a1c8a315fd231707,Johannsen_2010,PhysRevD.83.124015,Cardoso:2014rha,Rezzolla:2014mua,Konoplya:2016jvv,Berti:2015itd,DelPiano:2025ykr}) and usually depend on a number of fit parameters to describe the deviation from the classical solution. These parameters are fixed by adapting the metric either to fit (experimentally measured) observables (see \emph{e.g.} \cite{Cadoni:2022vsn}) or to approximate (more complicated) solutions stemming from theoretical models. In the recent article \cite{DelPiano:2025ykr} several such approaches along with the \emph{Effective Metric Description} (EMD) \cite{DelPiano:2023fiw,DelPiano:2024gvw,DelPiano:2024nrl}, have been compared, notably in view of the number of fit parameters required to achieve a predefined accuracy for certain observables.  The EMD is a different effective approach, which describes deviations of the metric (relative to the Schwarzschild black hole) in terms of physically relevant quantities. Indeed, these metric deformations are written as functions of a physical quantity (\emph{e.g.} the spatial distance to the event horizon \cite{Binetti:2022xdi,DAlise:2023hls,DelPiano:2023fiw,DamiaPaciarini:2024ssf} (see \cite{Myszkowski:2025hdc} for an EMD based on a physical distance to a generic point in space-time), a curvature invariant \cite{DelPiano:2024gvw} or the proper time of a free-falling observer to reach the horizon \cite{DamiaPaciarini:2025xjc,SH}). These can more directly be linked to effective physics of the black hole (see \emph{e.g.} \cite{SH}) or dual descriptions of gravity (\emph{e.g.} through AdS-CFT duality \cite{Maldacena:1997re}) than mere fit-parameters. However, since such physical quantities generally depend on the metric itself, the EMD requires to solve an intricate, highly non-linear differential equation in a self-consistent fashion. This is achieved by developing the metric deformations in power series of the physical quantity and providing recursive relations to fix the expansion coefficients unambiguously. The latter indeed carry physical significance as they can directly be related to physical measurements by local observers in the vicinity of the black hole \cite{SH}. Due to convergence of the power series, this approach is typically confined to a limited region in space time (\emph{e.g.} the close vicinity of the event horizon). The range of applicability can, however, be extended \cite{DelPiano:2024nrl} by replacing the series expansions by Padé approximants, allowing to (approximately) compute observables, such as the black hole shadow, at a larger distance to the horizon.

Inspired by the EMD, we propose a new kind of effective description that is also based on physical quantities, but directly valid everywhere outside of the horizon: while the EMD expresses metric deformations as expansions in a physical quantity that is small in a certain region of space-time (\emph{i.e.} it develops observables \emph{close} to a certain point, such as the horizon), in this work we shall organise this expansion in terms of a physical quantity that is small because the geometry under consideration is 'close' to the benchmark one. Such a quantity should be observer independent and identically vanishing in the case of the Schwarzschild space-time (at least outside of the horizon). Furthermore, ideally, it should be easily relatable to the physics in the vicinity of the black hole, thus facilitating the connection to theories of (quantum) gravity beyond GR. We choose to this end the eigenvalues of the Einstein tensor. More specifically, while a generic spherically symmetric and static black hole has two functionally independent eigenvalues, we shall for simplicity restrict ourselves to deformed black holes which only have one independent eigenvalue.\footnote{This assumption leads to a further simplification of the possible metrics for the black hole, thus simplifying the discussion in order to better highlight the idea of our approach. We plan to generalise our approach in the future by relaxing this assumption.} After suitable normalisation by the mass of the black hole, we shall denote the latter $\eps$, which vanishes identically outside of the horizon for the Schwarzschild black hole. In order to describe deformed black holes, we shall consider $\eps$ a given function of the space-time position, thereby fully determining the geometry and all observables derived from it (albeit in a highly complicated fashion). Assuming that $\eps$ is sufficiently regular and small (as well as all its derivatives) everywhere outside of the black hole, we provide approximations of not only the geometry itself, but also numerous observables in the form of linearised functionals of $\eps$.\footnote{Since we restrict ourselves to the space-time outside of the horizon of the black hole, some observables also depend on an additional small constant, which can for example be chosen to be the position of the horizon relative to the Schwarzschild geometry.}

The choice of $\eps$ as a quantity to describe black hole deformations beyond GR also furnishes a connection to (effective) physics outside of the black hole. Indeed, for a given $\eps$, the black hole geometry can be read as solution of an Einstein equation with a type I (in the classification of Hawking-Ellis \cite{Hawking:1973uf}) energy momentum tensor ${T^\mu}_\nu$ as external source, whose eigenvalue is $\eps$. This description is, however, not fundamental but only an effective interpretation of the black hole solution. Nevertheless, ${T^\mu}_\nu$ can be seen to arise from external physical systems, that source the black hole geometry, which may further be linked to other, more fundamental descriptions (\emph{e.g.} a theory of quantum gravity). The approach described in this work, therefore provides an intermediate step between (measured) observables on one side and fundamental theories of (quantum) gravity on the other. 

Our expressions for such observables are similar to the linear response formalism for general physical systems (see \emph{e.g.} \cite{Kubo,Green,Streda_1982}). Indeed, given a system in a stable state, observables of the same system under the influence of a small external perturbation, can be understood as functionals of the latter. When expanding this functional in powers of this deformation, often times already the first (linear) correction is sufficient for a satisfactory description of the perturbed system. Moreover, in many cases, the coefficient function of this leading term can be expressed purely in terms of the unperturbed system. In our approach, the stable, unperturbed sate of the system corresponds to the Schwarzschild geometry, while we interpret $\eps$ (and the constant mentioned above) as a small external perturbation. In this paper, we then provide explicit expressions for the leading linear response of the black hole to this perturbation for a variety of observables. The coefficients of this linear response are generally functions of space-time and independent of $\eps$ and thus universal. They are therefore expected to encode model-independent features of physics beyond GR. However, we stress once more that our treatment of $\eps$ as an \emph{external} perturbation is what renders our approach effective: in a truly fundamental theory of gravity beyond GR, $\eps$, as the eigenvalue of a suitable energy momentum tensor, would not be a mere external source for the metric, but would rather be itself the result of the interaction of matter with gravity (and thus not completely decoupled from the latter). Nevertheless, as we shall demonstrate in this work, the above mentioned systematic separation of model-dependent and universal features of observables, allows our formalism to extract useful local information of black hole geometries.

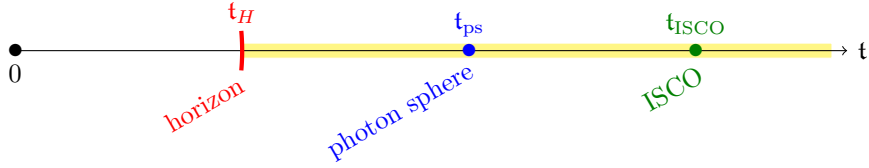
\begin{figure}[h]
\begin{center}
\begin{tikzpicture}
\draw[white,fill=white!40!yellow] (3,-0.1) -- (10.8,-0.1) -- (10.8,0.1) -- (3,0.1) -- (3,-0.1);
\node at (0,0) {$\bullet$};
\node at (0,-0.3) {\footnotesize $0$};
\draw[->] (0,0) -- (11,0);
\node at (11.2,0) {\footnotesize $\tco$};
\draw [domain=-5:5, ultra thick,red] plot ({3*cos(\x)}, {3*sin(\x)});
\node[red] at (3,0.5) {\footnotesize $\tco_H$};
\node[rotate=30,red] at (2.5,-0.7) {\footnotesize horizon}; 
\node[blue] at (6,0) {$\bullet$};
\node[blue] at (6,0.35) {\footnotesize $\tco_{\text{ps}}$};
\node[rotate=30,blue] at (5.1,-0.8) {\footnotesize photon sphere}; 
\node[green!50!black] at (9,0) {$\bullet$};
\node[green!50!black] at (9,0.35) {\footnotesize $\tco_{\text{ISCO}}$};
\node[rotate=30,green!50!black] at (8.7,-0.5) {\footnotesize ISCO}; 
\end{tikzpicture}
\end{center}
\caption{Schematic overview over important space-time positions at which observables discussed in this work provide information about the function $\eps$. For concreteness, we only consider the yellow shaded region, representing the exterior of the black hole (\emph{i.e.} outside of the event horizon) and parametrise space-time points (up to rotational symmetry) through the suitably normalised proper time $\tco$ of a radially in-falling massive observer to reach the horizon. The latter (shown in red in the figure) is located at $\tco_H$ where we obtain information through the surface gravity (\emph{i.e.} observable \emph{(i)}). Close to the photon sphere at $\tco_{\text{ps}}$ (shown in blue), we obtain information through the observables \emph{(ii)} and \emph{(iv)}, \emph{i.e.} the geodesic motion of massless particles and QNMs of massless scalar fields. Finally, at the ISCO (shown in green), which approximately marks the inner rim of a possible accretion disk of the black hole, we can extract information from the geodesic motion of massive test particles, \emph{i.e.} observables \emph{(iii)}.}
\label{Fig:LampPost}
\end{figure}

\noindent
Concretely, we shall discuss the following observables in this work
\begin{itemize}
\item[\emph{(i)}] the surface gravity $\kappa$, which can be related to the Hawking temperature of the black hole
\item[\emph{(ii)}] characteristic quantities of the geodesic motion of massless test particles in the vicinity of the black hole, notably the black hole shadow (which was also previously discussed in \cite{DelPiano:2024nrl} in the framework of the EMD), the frequency of photons on a circular orbit around the black hole and the Lyapunov exponent \cite{Cardoso:2008bp} (which provides information about the stability of the photon sphere)
\item[\emph{(iii)}] characteristic quantities of the geodesic motion of massive test particles in the vicinity of the black hole, notably the position of the innermost stable circular orbit (ISCO)
\item[\emph{(iv)}] the spectrum of quasi-normal modes (QNMs) of massless scalar fields in the background of the black hole geometry  
\end{itemize}
Each of these observables encodes information about the black hole geometry at a particular region in space-time. More concretely, as we shall explain, each of these observables depends on $\eps$ (and/or its derivatives) evaluated at a particular distance to the black hole horizon or on (nested) integrals of $\eps$ from the horizon to this distance. Therefore, using the observables \emph{(i)} -- \emph{(iv)} as input, furnishes (approximate) integro-differential equations for $\eps$ as a function of space-time (see Figure~\ref{Fig:LampPost} for a schematic overview). Our approach therefore provides a direct relation between experimentally accessible observables of the black hole and a quantity $\eps$, which can (in principle) be connected to an effective description of the physics around the black hole. Notably, the linear response framework allows to clearly distinguish features that are universal (\emph{i.e.} which are uniquely a property of the observable under consideration) and features that depend on $\eps$ and are thus related to new physics beyond the vacuum GR benchmark. We also remark that our approach is independent of the chosen parametrisation (\emph{i.e.} the choice of coordinates in which we specify $\eps$ as a function of the space-time position). For concreteness, here we use as coordinate the (suitably normalised) proper time $\tco$ of a massive, radially in-falling test particle to reach the horizon to label space-time points (up to rotational symmetry) and provide explicit relations how to transform to other parametrisations. To leading order in $\eps$, these only depend on relations that are accessible in the classical Schwarzschild geometry. Finally, we illustrate our findings by using three metric deformation functions, that have been used in models of black hole geometries beyond GR.

This paper is organised as follows: in Section~\ref{Sect:LinearResponse}, we develop the linear response approach in detail for a generic observable. We discuss how to adapt the result for different parametrisations of space-time and introduce three concrete examples of deformed black hole geometries, which serve to illustrate our results in subsequent Sections. In Section~\ref{Sect:GeodesicMotion}, we discuss observables related to the geodesic motion of massless and massive particles around the black hole. These include notably the position of the photon sphere and the ISCO as well as critical frequencies of test-particles orbiting the black hole at these respective loci. In Section~\ref{Sect:QNM}, we analyse the frequencies of scalar Quasi-normal Modes from the perspective of the linear response approach. We particularly show, how the full spectrum allows to deduce a non-trivial functional relation for the deformation function $\eps$ in the vicinity of the photon sphere. Section~\ref{Sect:Conclusions} contains our conclusions and outlook for future work. This paper is complemented by 2 Appendices, which contain additional details on some of the observables considered in the main body of this work and further technical aspects related to the concrete examples that are used to illustrate our theoretical results. 
%%%%%%%%%%%%%%%%%%%%%%%%%%%%%
\section{Deformation Function and Linear Response}\label{Sect:LinearResponse}
\subsection{General Idea and Linear Response}
In this work, we shall develop a framework to describe deformations of a static, spherically symmetric black hole beyond a solution of General Relativity (GR), that resembles linear response theory. For a general physical system, linear response theory requires three ingredients:
\begin{itemize}
\item[\emph{(i)}] a \emph{ground state}, as a (stable) solution of the system in the absence of perturbations: in the case at hand, this is the Schwarzschild space-time as a static, spherically symmetric (non-extremal) black hole solution in GR. We shall denote the mass of this black hole by $M$ and we shall only consider the space-time outside of the event horizon.
\item[\emph{(ii)}] a small \emph{deformation} $\mathfrak{F}$ of the system: we shall consider a slightly deformed black hole geometry that is not longer an exact solution of GR, but rather of some theory beyond it, \emph{e.g.} a theory of quantum gravity, or a theory that introduces higher order terms beyond the Einstein-Hilbert action. We shall assume, however, that the geometry is still static and spherically symmetric. Furthermore, (inspired by the EMD framework developed in \cite{DelPiano:2023fiw,DelPiano:2024gvw,DelPiano:2024nrl}), we shall describe this deformation by an observer independent, invariant function that is defined at every point in (a region $\mathcal{S}$ of) space-time, outside of the event horizon. Due to the symmetries of the space-time, the latter can be parametrised by a single suitable parameter $\lambda$.  
\item[\emph{(iii)}] an observable $\mathbb{O}[\mathfrak{F}]$ in the deformed theory, which is determined as a linear functional of the deformation, \emph{i.e.} schematically
\begin{align}
\mathbb{O}[\mathfrak{F}](\lambda)=\mathbb{O}[0](\lambda)+\int_{\mathcal{S}} \frac{\delta \mathbb{O}[\mathfrak{F}](\lambda)}{\delta \mathfrak{F}(\lambda')}\bigg|_{\mathfrak{F}=0}\,\mathfrak{F}(\lambda')\,d\lambda'+\mathcal{O}(\mathfrak{F}^2)\,,\label{LinearResponseForm}
\end{align}
where $\mathbb{O}[0](\lambda)$ is the observable in the absence of the deformation, while the second term on the right hand side, which is of order $\mathcal{O}(\mathfrak{F})$, represents the linear response of the system to the deformation. In the case at hand, $\mathbb{O}[0](\lambda)$ shall denote an invariant in the Schwarzschild geometry, which can be computed exactly using GR. We shall provide an approximate expression for certain observables in the deformed geometry, that resembles (\ref{LinearResponseForm}), namely as expressions that are linear in the deformation. These observables are related to physics around the black hole, \emph{e.g.} related to the movement of (test) particles around the black hole, or the spectrum of quasi-normal-modes (QNM) of scalar fields. 
\end{itemize}
The full functional $\mathbb{O}[\mathfrak{F}](\lambda)$ is in principle fixed by a consistent theory beyond GR, \emph{e.g.} a theory of quantum gravity. In this work, we shall consider the full form of this functional to be unknown and instead we shall develop an effective framework to capture the linear deformations (\emph{i.e.} the second term on the right hand side of (\ref{LinearResponseForm})). Our principal assumption is that, even in presence of $\mathfrak{F}$, the physics in the vicinity of the black hole is still captured geometrically.  In other words, the space-time outside of the horizon is described by a curved manifold $\mathcal{M}$, such that for example particle motion follows geodesics and the spectrum of scalar QNMs can be accessed by studying the Klein-Gordon equation on $\mathcal{M}$.

\subsection{Energy Momentum Tensor and Effective Framework}
We shall describe the deformation beyond the Schwarzschild geometry (both of the geometry itself, as well as for observables) in terms of a function that is unambiguously defined in every point of space-time and a constant, which acts as additional boundary condition for certain observables. It was argued in \cite{SH} that the near horizon metric can not only be reconstructed from telemetric data collected outside of the black hole, but also from the invariant eigenvalues of the Einstein tensor. In the current work, we shall assume a simplified geometry, in which only one of these eigenvalues is functionally independent and which shall act as deformation function. That is, we shall consider this eigenvalue (as a function of the position in the space-time region $\mathcal{S}$) to be a given input (\emph{i.e.} an external deformation in the language of the previous Subsection), from which we shall develop approximations of the geometry and observables to leading order. For certain observables, we shall require an additional constant, which we shall choose as (the square-root of) the area of the event horizon, which can be linked to leading order to other invariant quantities.

Parametrising black hole deformations through external functions that are identified with eigenvalues of the Einstein tensor, offers the possibility of connecting with other effective physics. Indeed, we may interpret the black hole geometry as a solution of the Einstein equation 
\begin{align}
R_{\mu\nu}-\frac{1}{2}\,g_{\mu\nu}\,R=T_{\mu\nu}\,,
\end{align}
where $g_{\mu\nu}$ is the metric, $R_{\mu\nu}$ and $R$ the Ricci-tensor and Ricci-scalar respectively and $T_{\mu\nu}$ an energy momentum tensor\footnote{To simplify numerical coefficients, we use a convention in this work, where $T_{\mu\nu}$ is rescaled by a factor of $8\pi$.}, which we interpret as external source (and which is characterised by the above-mentioned eigenvalues). To be more concrete, in the classification of Hawking-Ellis \cite{Hawking:1973uf}, ${T^\mu}_\nu$ is of type I, with one time-like eigenvalue, which we shall denote $\epsilon$. The latter is generically referred to as \emph{energy density}, while the remaining eigenvalues of ${T^\mu}_\nu$ represent parallel ($p_{||}$) and perpendicular ($p_{\perp}$) pressure components with respect to the radial direction. The simplification of the geometry mentioned above consists in assuming $\epsilon=-p_{||}$, while (due to the symmetries of the problem) also $p_{\perp}$ is related to $\epsilon$ in a more complicated form (see \emph{e.g.} \cite{Hawking:1973uf} as well as \cite{DAlise:2023hls}). We have chosen $\epsilon$ to encode the deformations of the Schwarzschild metric, for three reasons
\begin{itemize}
\item $\epsilon$ is an observer independent (Lorentz) invariant
\item $\epsilon=0$ in the Schwarzschild geometry outside of the event horizon, such that $\epsilon$ is directly sensitive to any deformations of a spherically symmetric and static (vacuum) black hole, going beyond GR
\item The energy momentum tensor ${T^\mu}_\nu$ (and also its eigenvalues) can be linked directly to effective physics, \emph{e.g.} in the form of effective actions of matter fields coupled to gravity. For example, in \cite{Boonserm:2015aqa} a minimal representation of a generic type I energy momentum tensor in the form of physical fields was proposed. Basing our effective black holes on $\epsilon$ allows to make more direct contact to concrete effective theories going beyond GR.
\end{itemize}
In view of the second point, we shall assume $\epsilon$ to be a small, given function at every point outside of the event horizon of the black hole, to parametrise a deformation of the metric as we shall now explain.

%%%%%%%%%%%%%%%%%%%%%%%%%%%%%%%%%%%%%

\subsection{Linear Response of Observables}
\subsubsection{Metric}
In order to obtain an effective linear response for the black hole geometry, we shall first develop the metric. Concretely, due to the symmetries of the geometry and the assumption $\epsilon=-p_{||}$, we shall consider
\begin{equation}
\parbox{6cm}{
\begin{align}
ds^2&=g_{\mu\nu}\,dX^\mu\,dX^\nu\nonumber\\
&=-f(z)\,dt^2+\frac{dz^2}{f(z)}+z^2\,d\theta^2+z^2\,\sin^2\theta\,d\varphi^2\,,\nonumber
\end{align}
}
\hspace{1cm}\text{with}\hspace{1cm}\begin{array}{l}X^\mu\in\{t,z,\theta,\varphi\}^\mu\,, \\[6pt]f(z)=1-\frac{\Phi(\tau)}{z}\,.\end{array}\label{FormMetric}
\end{equation}
This metric is characterised by a single\footnote{The assumed symmetries fix the geometry up to two independent functions of the radial coordinate (see \emph{e.g.} \cite{Carroll:2004st}). As mentioned before, in this work, for simplicity, we restrict ourselves to geometries of the form (\ref{FormMetric}).} function $f(z)$ of the radial coordinate. In the undeformed case (which serves as the ground state for the linear response approach), (\ref{FormMetric}) represents the Schwarzschild geometry, for which $f(z)=:f_S(z)=1-\frac{2M}{z}$. The deformed case is characterised by the function $\Phi$ in (\ref{FormMetric}), which we choose to write as a non-trivial function of the proper time $\tau$ for a massive test-particle, following a radial geodesic, to reach the horizon located at $\zh$ (see Figure~\ref{Fig:Infalling} below for a schematic drawing)
\begin{align}
&\tau(z)=\int_{z_H}^z\frac{du}{\sqrt{|1-f(u)|}}\,,\label{DefTau}
\end{align}
which is understood as a function of the radial coordinate $z$. As we shall discuss in Section~\ref{Sect:Parametrisation}, while the choice of $\tau$ as parametrisation is ambiguous, it has no conceptual impact on our formalism. 

From the Einstein tensor associated with (\ref{FormMetric}), we obtain the eigenvalue $\epsilon$ mentioned above\footnote{For brevity, we shall refer to this quantity as \emph{energy density} in the following.}
\begin{align}
&\epsilon=\frac{1-f(z)-z\,f'(z)}{z^2}\,,&&\text{with} &&f'(z)=\frac{df}{dz}\,.\label{DefTauR}
\end{align}
In order to further simplify the discussion, we use dimensionless quantities in this work. Denoting $M$ the mass of the black hole, we introduce
\begin{align}
&r=\frac{z}{M}\,,&&\tco:=\frac{\tau}{M}\,,&&\eps(\tco):=M^2\,\epsilon(\tau)\,,&&\fun(\tco):=\frac{\Phi(\tau)}{M}\,.\label{DimensionlessQuant}
\end{align}
Similarly, instead of $z_H$ in (\ref{DefTau}) (which marks the position of the event horizon and is a zero of the function $f$), we shall also use the dimensionless quantity $\rh=\frac{z_H}{M}$, which can be related to the area $A_{\text{H}}$ (in units of $M^2$) of the horizon through (see \emph{e.g.} \cite{Carroll:2004st}, eq.~(6.107))
\begin{align}
A_{\text{H}}=\rh^2\,\int_0^{2\pi}d\varphi\int_0^\pi\sin\theta\,d\theta =4\pi\,\rh^2\,.
\end{align}
Since $\rh\to 2\neq 0$ for $\epsilon\to 0$ (\emph{i.e.} in the Schwarzschild geometry), we shall instead define the following dimensionless small constant
\begin{align}
\con:=\rh-2\,,
\end{align} 
which is required to describe the linear response of certain observables. As we shall discuss later on, other choices for $\con$ are possible, since $\rh$ can also be related to further invariants of the black hole, such as the value of the Kretschmann scalar at the horizon (see (\ref{rhKretschRel})).

\subsubsection{Linear Response}\label{Sect:LinRespsone}
In \cite{SH}, the following relation was found (in dimensionful notation)
\begin{align}
&\epsilon(\tau)=-p_{||}(\tau)\approx\frac{1}{M\, (3\tau+4M)}\,\frac{d}{d\tau}\Phi(\tau)\,,\label{RelImportEpsPhi}\\
&z(\tau)\approx\frac{1}{\left(8+\frac{6\tau}{M}\right)^{1/3}}\left(2(Mr_H+\tau)+\frac{1}{2M}\int_0^\tau dt\,\Phi(t)\right)\,.\label{EffectiveProbePosition}
\end{align}
where the $\approx$ indicates an approximate relation that holds true for small deformation functions of the metric.\footnote{More precisely, in \cite{SH}, the geometry was studied in the EMD framework, where the metric function $\Phi$ was written in a series expansion in powers of $\tau$ (see (\ref{SeriesExpansion})), with small coefficients $\{v_n\}$. The relations (\ref{RelImportEpsPhi}) and (\ref{EffectiveProbePosition}) hold up to corrections of order $\mathcal{O}(v_n^2)$. For a brief review of this approach, see Appendix~\ref{App:SeriesExpansion}.} The relation (\ref{RelImportEpsPhi}) can be inverted to approximate $\Phi$ as a functional of $\epsilon$, which (along with (\ref{EffectiveProbePosition})) reads in dimensionless notation
%\begin{align}
%\Phi(\tau)\sim\rh+M \int_0^\tau\,d\tau'(3\tau'+4M)\, \epsilon(\tau')\,.
%\end{align}
%In the dimensionless notation introduced in the previous Subsubsection, we can write
\begin{align}
&\fun(\tco)=\rh+\int_0^{\tco}\,d\tcop\,\left(3\tcop+4\right)\,\eps(\tcop)+\mathcal{O}(\con^2,\eps^2,\con\eps)\,,\label{DeformationFunction}\\
&r(\tco)=\frac{1}{(8+6\tco)^{1/3}}\left[2\rh\,\left(1+\frac{\tco}{4}\right)+2\tco+\frac{1}{2}\int_0^{\tco}d\tcop\int_0^{\tcop}d\tcop' (3\tcop' +4)\,\eps(\tcop')\right]+\mathcal{O}(\con^2,\eps^2,\con\eps)\,.\label{RadialCoord}
\end{align}
In the following we use the notation $\sim$ to denote relations that are correct up to non-linear terms in the deformation function $\eps$ and $\con$. Using (\ref{RadialCoord}) one can directly check that
\begin{align}
&\frac{1-f-r\tfrac{df}{dr}}{r^2}\sim \eps\,,&&\text{and} &&\frac{dr}{dt}\sim \sqrt{1-f}\,,
\end{align}
which are compatible with (\ref{DefTau}) and (\ref{DefTauR}) respectively. Notice that these relations do not rely on any series expansions as in \cite{SH} and are thus not limited by any radius of convergence. This means that (\ref{DeformationFunction}) and (\ref{RadialCoord}) are valid anywhere outside the event horizon provided that $\con$ and $\eps$ (as well as its derivatives and nested integrals) remain small. The linearised relations (\ref{DeformationFunction}) and (\ref{RadialCoord}) further allow to write for the metric function $f$ appearing in (\ref{FormMetric}) 
\begin{align}
f(\tco)&\sim 1-\frac{8\left(4+\left(1+\rh\right)\tco\right)}{\left(8+6\tco\right)^{5/3}}-\frac{2}{\left(8+6\tco\right)^{2/3}}\int_0^\tco\,d\tcop(3\tcop+4)\, \eps(\tcop)\nonumber\\
&\hspace{1cm}+\frac{4}{\left(8+6\tco\right)^{5/3}}\int_0^\tco d\tcop\int_0^{\tcop}\,d\tcop' (3\tcop'+4)\,\eps(\tcop')\,.\label{FormFcomp}
\end{align}
Since $f$ encodes the metric of the deformed space-time, general invariants (\emph{i.e.} Lorentz scalars) $\mathbb{O}[\con,\eps]$ can be computed from it in a similar linearised fashion. The form (\ref{DeformationFunction}) and (\ref{RadialCoord}) thus suggests that we can write in general
\begin{align}
&\mathbb{O}[\con,\eps](\tco)\sim\mathbb{O}[0,0](\tco)+\coff{\mathbb{O}}{\con}(\tco)\,\con+\sum_{n\in\mathbb{Z}}\coff{\mathbb{O}}{n}(\tco)\,\mathcal{I}_n[\eps](\tco)\,,\label{LinearResponse}
\end{align}
with the following \emph{momenta}
\begin{align}
I_n[\eps](\tco)=\left\{\begin{array}{lcl}\frac{d^{|n|}\eps}{d\tco^{|n|}}(\tco) & \text{for} & n<0\,, \\[6pt] \eps(\tco) & \text{for} & n=0\,, \\[6pt](3\tco+4)^{1-n} \int_0^\tco d\tco_1\int_0^{\tco_1}d\tco_2\ldots\,\int_0^{\tco_{n-1}}d\tco_n\,(3\tco_n+4)\,\eps(\tco_n) & \text{for} & n>0\,.\end{array}\right.\label{Momenta}
\end{align}
We can also interpret (\ref{LinearResponse}) as a linear integro-differential operator acting on $\eps$. However, for (\ref{LinearResponse}) to represent a useful approximation of $\mathbb{O}$, we assume that all contributing momenta are small (in the relevant region of space-time). In Section~\ref{Sect:ExamplesMetrics} we shall discuss a number of examples, in which this is indeed the case. Furthermore, we implicitly assume that $\eps$ is monotonic such that $0<f<1$ (at least in the relevant region of space-time) and $\tco$ is well defined. Finally, as we shall see, for most invariants (which we shall call observables) that we shall study in this paper, only finitely many $\coff{\mathbb{O}}{n}(\tco)$ are different from zero, such that the summation in (\ref{LinearResponse}) is in general over finitely many terms. Indeed, we find for the metric deformation function $\fun$ in~(\ref{DeformationFunction})
\begin{align}
&\mathcal{F}(\tco)\big|_{\con=0,\eps=0}=2\,,&&\coff{\mathcal{F}}{\con}(\tco)=1\,,&&\coff{\mathcal{F}}{1}(\tco)=1\,,&&\coff{\mathcal{F}}{n}=0\hspace{0.3cm}\forall n\neq 1\,,\label{LinRespPos}
\end{align}
and for the radial variable $r$ in (\ref{RadialCoord}) 
\begin{align}
&r(\tco)\big|_{\con=0,\eps=0}=\tfrac{1}{2}(8+6\tco)^{2/3}\,,&&\coff{r}{\con}(\tco)=\frac{2+\tfrac{\tco}{2}}{(8+6\tco)^{1/3}}\,,\nonumber\\
&\coff{r}{2}(\tco)=\tfrac{1}{4}(8+6\tco)^{2/3}\,,&&\coff{r}{n}(\tco)=0\hspace{0.3cm}\forall n\neq 2\,,\label{rLeadingForms}
\end{align}
and the metric function $f$ in (\ref{FormFcomp}) 
\begin{align}
&f(\tco)\big|_{\con=0,\eps=0}=1-\frac{4}{(8+6\tco)^{2/3}}\,,&&\coff{f}{\con}(\tco)=-\frac{8\tco}{(8+6\tco)^{5/3}}\,,\nonumber\\
&\coff{f}{1}(\tco)=-\frac{2}{(8+6\tco)^{2/3}}\,,&&\coff{f}{2}(\tco)=\frac{2}{(8+6\tco)^{2/3}}\,,&&\coff{f}{n}(\tco)=0\hspace{0.3cm}\forall n\notin\{1,2\}\,.
\end{align}
The form (\ref{LinearResponse}) strongly resembles (\ref{LinearResponseForm}): the term $\mathbb{O}[0,0](\tco)$ denotes the value of the observable in the absence of any deformations and can thus be exactly calculated within GR. The terms proportional to $\con$ and $\mathcal{I}_n$ encode the linear response of the black hole geometry and in the following we shall develop the coefficient functions $\coff{\mathbb{O}}{\con}$ and $\{\coff{\mathbb{O}}{n}\}_{n\in\mathbb{Z}}$ for a number of invariants. To keep the notation light, we shall in the following omit terms of order $\mathcal{O}(\con^2,\eps^2,\con\eps)$ and denote relations that hold within the linear response by $\sim$. To further lighten the notation, we shall refrain from writing coefficients $\coff{\mathbb{O}}{\con,n}$ that are identically vanishing and in certain cases we shall use the shorthand notation $\sh(\tco):=8+6\tco$.

%%%%%%%%%%%%%%%%%%%%%%%%%%%%%%%%%%
\subsection{Parametrisation}\label{Sect:Parametrisation}
Before moving on to discuss concrete invariants and observables, we shall first address the parametrisation of the deformation function $\eps$ and, as a consequence, also the invariants and (physical) observables computed from it. A basic assumption of our approach is that the physics outside of the black hole is described geometrically, \emph{i.e.} by representing space-time in the form of a curved manifold $\mathcal{M}$ (which can be captured by a metric, such as the one in (\ref{FormMetric})). In order to describe deformations, we need to assign a value $\eps$ at every point $\mathcal{P}\in\mathcal{M}$, thus requiring a

\begin{wrapfigure}{l}{0.45\textwidth}
\vspace{-0.75cm}
\begin{center}
\includegraphics[width=7.5cm,trim={0 4.5cm 0 0}]{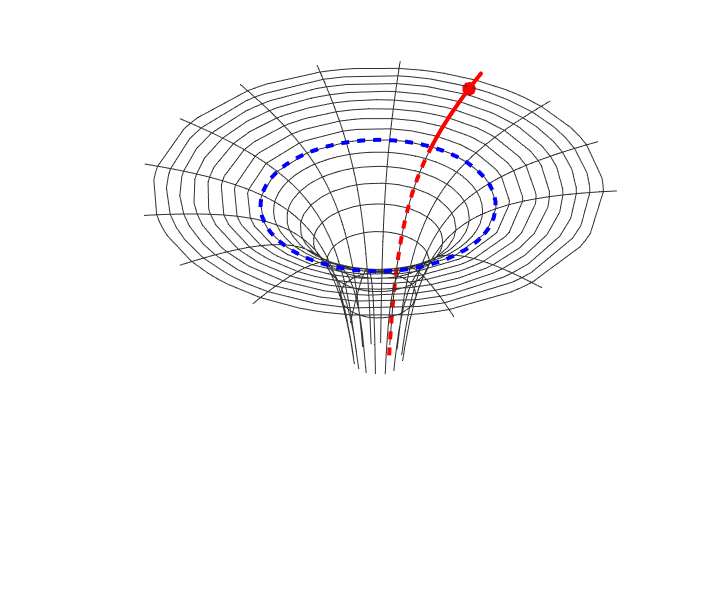}
\end{center}
\vspace{-0.5cm}
\caption{Schematic drawing of a (massive) particle following a radial, infalling geodesic (red line) in the black hole geometry. The dashed blue line represents the event horizon.}
\label{Fig:Infalling}
\vspace{-0.5cm}
\end{wrapfigure}

\noindent
parametrisation of (a patch of) $\mathcal{M}$. A priori, any such parametrisation is viable, however, in order to describe physically meaningful invariants (\emph{i.e.} observables) within our framework, we require it to be observer independent. Moreover, we prefer a choice of coordinates that can be introduced in a simple invariant fashion, that is related to (physically) measurable quantities. Following \cite{SH} we have chosen as such a quantity the suitably normalised proper time $\tco\geq 0$ of a massive, free-falling observer (\emph{i.e.} on a radial trajectory) to reach the event horizon. For a schematic drawing, see Figure~\ref{Fig:Infalling}. Up to spherical symmetry, this proper time uniquely determines a point in space-time and can conceptually be introduced without the need to define a metric. For this reason, we have chosen to parametrise $\eps$ and all invariants $\mathbb{O}$ in terms of $\tco$ in the previous Subsection. 
%\begin{figure}[h]
%\begin{center}
%\includegraphics[width=10cm,trim={0 4.5cm 0 0}]{InfallingSchematic.png}
%\end{center}
%\caption{Schematic drawing of a (massive) particle following a radial, infalling geodesic (red line) in the black hole geometry. The dashed blue line represents the event horizon.}
%\label{Fig:Infalling}
%\end{figure}
However, $\tco$ is not the only possible choice of parametrisation. For exemple, the coordinate $r$ appearing in (\ref{FormMetric}) would also be a viable parametrisation for $\mathcal{P}\in\mathcal{M}$.\footnote{A possible invariant definition would be the (suitably normalised) area of a sphere, concentric with the center of the black hole, on which the point is localised. Although this definition seems more complicated to realise.} Another possibility, which is particularly suitable for static black holes considered here, is the spatial distance to the event horizon,% In units of the black hole mass, this distance can be defined with the help of the metric (\ref{FormMetric})
%\begin{align}
%\rho(r)=\int_{\rh}^{r}\frac{dr'}{\sqrt{|f(r')|}}\,,
%\end{align}
which was advocated in \cite{DelPiano:2023fiw,DelPiano:2024gvw,DelPiano:2024nrl} as a universal way to parametrise deformations of the Schwarzschild metric in the form of EMDs. It is therefore important to discuss how a change of parametrisation modifies the linear response (\ref{FormFcomp}).
%%%%%%%%%%%%%%%%
\subsubsection{General Change of Parametrisation}\label{Sect:VariableChange}
\begin{figure}[h]
\begin{center}
\begin{tikzpicture}
\draw[white,fill=white!40!yellow] (3,-0.1) -- (8.8,-0.1) -- (8.8,0.1) -- (3,0.1) -- (3,-0.1);
\node at (0,0) {$\bullet$};
\node at (0,-0.35) {\footnotesize origin};
\draw[->] (0,0) -- (9,0);
%\node at (11.3,0) {\footnotesize $\tco$, $x$};
\draw [domain=-5:5, ultra thick,red] plot ({3*cos(\x)}, {3*sin(\x)});
\node[red] at (3,0.5) {\footnotesize $\tco_H$, $x_H$};
\node[rotate=30,red] at (2.5,-0.7) {\footnotesize horizon}; 
\node at (6,0) {$\bullet$};
\node at (6,0.35) {\footnotesize $\eps(\tco)=\eps_x(x)$};
\node at (6,-0.4) {\footnotesize $\mathcal{P}$}; 
\node[yellow!90!black] at (8.5,-0.4) {\footnotesize $\mathcal{M}$};
\end{tikzpicture}
\caption{Schematic overview of two parametrisations: the energy density is (up to symmetries of the geometry) uniquely defined at each point. We shall consider two parametrisations in terms of $\tco$ (as in (\ref{DimensionlessQuant}) and (\ref{DefTau}) and a second parameter $x$, which are related through a bijection. For further clarity, we shall call $\eps_x$ the energy density parametrised by $x$, such that for a point $\mathcal{P}$ with coordinates $\tco$ and $x$ respectively, we have $\eps(\tco)=\eps_x(x)$.}
\label{Fig:Re-parametrisation}
\end{center}
\end{figure}
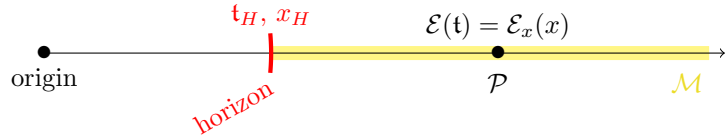

Let $\mathcal{P}\in\mathcal{M}$ be a point in space-time, which (up to spherical symmetry) is equivalently parametrised by $\tco$ (as in (\ref{DimensionlessQuant}) and (\ref{DefTau})) and some other viable parameter $x$. We assume that they are related through a bijection everywhere on $\mathcal{M}$. Furthermore, let us denote $\eps_{x}(x)$ the energy density parametrised by $x$, which agrees with $\eps(\tco)$ at $\mathcal{P}$, as schematically shown in Figure~\ref{Fig:Re-parametrisation}. We may then equally write the linear response for an observable $\mathbb{O}$ in the form

\begin{align}
&\mathbb{O}_{\xco}[\con,\eps_\xco](x)\sim\mathbb{O}_{\xco}[0,0](\xco)+\coff{\mathbb{O}}{x,\con}(\xco)\,\con+\sum_{n\in\mathbb{Z}}\coff{\mathbb{O}}{\xco,n}(\tco)\,\mathcal{I}_{\xco,n}[\eps_\xco](\xco)\,,\label{LinearResponseX}
\end{align}
For clarity, we have added subscripts $\xco$ to all quantities involved to indicate that they are a priori different functions than in (\ref{LinearResponse}). Indeed, as is for example evident from (\ref{RadialCoord}), already the change of the parameter itself may involve contributions from $\con$ and $\eps$: generalising (\ref{rLeadingForms}) we may write
\begin{align}
\xco(\tco)\sim \xco|_{\con=0,\eps=0}(\tco)+\coff{x}{\con}(\tco)\,\con+\sum_{n\in\mathbb{Z}}\coff{x}{n}(\tco)\,\mathcal{I}_n[\eps](\tco)\,,\label{GenFormXVar}
\end{align}
along with the linear response for the inverse
\begin{align}
\tco(\xco)\sim \tco|_{\con=0,\eps_\xco=0}(\xco)+\coff{\tco}{\xco,\con}(\xco)\,\con+\sum_{n\in\mathbb{Z}}\coff{\tco}{\xco,n}(\tco)\,\mathcal{I}_{\xco,n}[\eps_\xco](\xco)\,.\label{TxForm}
\end{align}
Here the leading terms $\xco|_{\con=0,\eps=0}(\tco)$ and $\tco|_{\con=0,\eps_\xco=0}(\xco)$ are determined by the classical change of parametrisation, \emph{i.e.} they can (at least locally) be explicitly computed within the undeformed Schwarzschild geometry. Similarly, the momenta $\mathcal{I}_{\xco,n}[\eps_\xco]$ in (\ref{GenFormXVar}) can\footnote{Alternative definitions would at leading order correspond to a linear redefinition of the momenta, with coefficients that are functions of $\xco|_{\con=0,\eps=0}(\tco)$ and $\tco|_{\con=0,\eps_\xco=0}(\xco)$.} be defined as in (\ref{Momenta}), with a change of variables for the derivatives and integrals based on the undeformed geometry,~\emph{i.e.}
\begin{align}
\frac{d}{d\tco}&\longrightarrow \frac{d\xco}{d\tco}\,\frac{d}{d\xco}=\frac{d\xco|_{\con=0,\eps=0}}{d\tco}(\tco|_{\con=0,\eps_x=0}(\xco))\,\frac{d}{d\xco}\,,\nonumber\\
\int_0^\tco\,d \tco'&\longrightarrow \int_{\xco_H}^{\xco|_{\con=0,\eps=0}(\tco)}\frac{d\tco|_{\con=0,\eps=0}}{d\xco}(\xco')\,d\xco'\,,\nonumber\\
3\tco+4&\longrightarrow 3\tco|_{\con=0,\eps_x=0}(\xco)+4\,.
\end{align}
where $\xco_H$ is the value of $\xco$ at the event horizon. 
%\begin{align}
%I_{x,n}[\eps_x](\xco)=\left\{\begin{array}{lcl}\frac{d^{n}\eps_{\xco}}{d\xco^n}(\xco) & \text{for} & n<0\,, \\[6pt] \eps_{\xco}(\xco) & \text{for} & n=0\,, \\[6pt]\frac{\int_{\xco_H}^\xco d\xco_1\ldots\,\int_{\xco_H}^{\xco_{n-1}}d\xco_n\,(3\tco|_{\con=0,\eps_x=0}(\xco_{n})+4)\,\eps_x(\xco_n)}{(3\tco|_{\con=0,\eps_x=0}(\xco)+4)^{n-1}} & \text{for} & n>0\,,\end{array}\right.\label{MomentaX}
%\end{align}
%where $\xco_H$ is the value of $\xco$ at the event horizon. 
Furthermore, the $\left\{\coff{x}{\con}(\tco),\left(\coff{x}{n}(\tco)\right)_{n\in\mathbb{Z}}\right\}$ in (\ref{GenFormXVar}) and the $\left\{\coff{\tco}{\xco,\con}(\xco),\left(\coff{\tco}{\xco,n}(\xco)\right)_{n\in\mathbb{Z}}\right\}$ in (\ref{LinearResponseX}) are suitable coefficient functions depending on $\tco$ and $\xco$ respectively, which can be computed within the linear response approach. These coefficients furthermore allow to relate $\mathbb{O}_{\xco}[\con,\eps](x)$ in (\ref{LinearResponseX}) to $\mathbb{O}[\con,\eps](\tco)$ in (\ref{LinearResponse}) for a given observable $\mathbb{O}$ at a point $\mathcal{P}$ on the manifold $\mathcal{M}$ that is equivalently parametrised by $\tco$ and $\xco$ (see Figure~\ref{Fig:Re-parametrisation}). In order to explain this, there are three different aspects that need to be considered:
\begin{itemize}
\item[\emph{(i)}] the {\bf leading contributions} $\mathbb{O}[0,0](\tco)$ and $\mathbb{O}_{\xco}[0,0](\xco)$: with (\ref{GenFormXVar}) and (\ref{TxForm}) we have
\begin{align}
\mathbb{O}_{\xco}[0,0](\xco)&\sim\mathbb{O}[0,0](\tco)+\frac{\partial \mathbb{O}_{\xco}[0,0]}{\partial \xco}\left(\xco|_{\con=0,\eps=0}(\tco)\right)\,\left(\coff{x}{\con}(\tco)\,\con+\sum_{n\in\mathbb{Z}}\coff{x}{n}(\tco)\,\mathcal{I}_n[\eps](\tco)\right)\,,\nonumber\\
\mathbb{O}[0,0](\tco)&\sim \mathbb{O}_{\xco}[0,0](\xco)+\frac{\partial \mathbb{O}[0,0]}{\partial \tco}\left(\tco|_{\con=0,\eps=0}(\xco)\right)\left(\coff{\tco}{\xco,\con}(\xco)\,\con+\sum_{n\in\mathbb{Z}}\coff{\tco}{\xco,n}(\tco)\,\mathcal{I}_{\xco,n}[\eps_\xco](\xco)\right)\,.\nonumber
\end{align}
where we have used $\mathbb{O}[0,0](\tco)=\mathbb{O}_{\xco}[0,0]\left(\xco|_{\con=0,\eps=0}(\tco)\right)$ and $\mathbb{O}_{\xco}[0,0](\xco)=\mathbb{O}[0,0](\tco|_{\con=0,\eps=0}(\xco))$, which represent the change of variables in the undeformed geometry.
 %%%%%%%%%%%%%%%%%%%%%%%%%
\item[\emph{(ii)}] the {\bf coefficient functions} $\left\{\coff{x}{\con}(\tco),\left(\coff{x}{n}(\tco)\right)_{n\in\mathbb{Z}}\right\}$ and the $\left\{\coff{\tco}{\xco,\con}(\xco),\left(\coff{\tco}{\xco,n}(\xco)\right)_{n\in\mathbb{Z}}\right\}$: since these coefficients multiply $\con$ or the momenta respectively, it is sufficient to perform a change of variables in the undeformed geometry, \emph{i.e.}
\begin{align}
&\coff{\tco}{\xco,\con}(\xco)=\coff{\tco}{\xco,\con}\left(\xco|_{\con=0,\eps=0}(\tco)\right)+\mathcal{O}(\con,\eps)\,,&&\text{and} &&\coff{\tco}{\xco,n}(\xco)= \coff{\tco}{\xco,n}\left(\xco|_{\con=0,\eps=0}(\tco)\right)+\mathcal{O}(\con,\eps)\,,\nonumber
\end{align}
and in the inverse case
\begin{align}
&\coff{\xco}{\con}(\tco)=\coff{\xco}{\con}\left(\tco|_{\con=0,\eps=0}(\xco)\right)+\mathcal{O}(\con,\eps)\,,&&\text{and} &&\coff{\xco}{n}(\tco)= \coff{\xco}{n}\left(\tco|_{\con=0,\eps=0}(\xco)\right)+\mathcal{O}(\con,\eps)\,.\nonumber
\end{align}
 %%%%%%%%%%%%%%%%%%%%%%%%%
\item[\emph{(iii)}] the {\bf momenta} $\mathcal{I}_{n}[\eps](\tco)$ and $\mathcal{I}_{\xco,n}[\eps_\xco](\xco)$: since the momenta are also of order 1 in $\con$ and $\eps$ it is sufficient to perform a change of variables based on the undeformed geometry, which is precisely the way in which the $\mathcal{I}_{\xco,n}[\eps_\xco](\xco)$ have been introduced. We therefore obtain
\begin{align}
&\mathcal{I}_{n}[\eps](\tco)\sim \mathcal{I}_{\xco,n}[\eps_\xco](\xco|_{\con=0,\eps_x=0}(\tco))\,,&&\text{and} &&\mathcal{I}_{\xco,n}[\eps_\xco](\xco)\sim \mathcal{I}_{n}[\eps](\tco|_{\con=0,\eps=0}(\xco))\,.
\end{align}
\end{itemize}
Taken these points together, we find that a change of parametrisation only leads to a shift in the coefficient functions $\left\{\coff{\mathbb{O}}{\con}(\tco),\left(\coff{\mathbb{O}}{n}(\tco)\right)_{n\in\mathbb{Z}}\right\}$ in a way that only depends on the observable of the undeformed geometry and the linear response coefficients of the change of variables (\ref{GenFormXVar}), concretely:
\begin{align}
\mathbb{O}_{\xco}[\con,\eps_\xco](x)&\sim\mathbb{O}[0,0](\tco)+\left(\coff{\mathbb{O}}{\con}(\tco)+\frac{\partial \mathbb{O}_{\xco}[0,0]}{\partial \xco}\left(\xco|_{\con=0,\eps=0}(\tco)\right)\,\coff{x}{\con}(\tco)\right)\,\con\nonumber\\
&\hspace{0.5cm}+\sum_{n\in\mathbb{Z}}\left(\coff{\mathbb{O}}{n}(\tco)+\frac{\partial \mathbb{O}_{\xco}[0,0]}{\partial \xco}\,\coff{x}{n}(\tco)\right)\,\mathcal{I}_{n}[\eps](\tco)\,.\label{LinearResponseXmod}
\end{align}
Similarly, we have the inverse relation
\begin{align}
\mathbb{O}[\con,\eps](\tco)&\sim\mathbb{O}_{\xco}[0,0](\xco)+\left(\coff{\mathbb{O}}{x,\con}(\xco)+\frac{\partial \mathbb{O}[0,0]}{\partial \tco}\left(\tco|_{\con=0,\eps=0}(\xco)\right)\,\coff{\tco}{\xco,\con}(\xco)\right)\,\con\nonumber\\
&\hspace{0.5cm}+\sum_{n\in\mathbb{Z}}\left(\coff{\mathbb{O}}{\xco,n}(\xco)+\frac{\partial \mathbb{O}[0,0]}{\partial \tco}\left(\tco|_{\con=0,\eps=0}(\xco)\right)\coff{\tco}{\xco,n}(\xco)\right)\,\mathcal{I}_{\xco,n}[\eps_\xco](\xco)\,.\label{LinearResponseInverseXmod}
\end{align}

\subsubsection{Example: Radial Coordinate}
As a concrete example, we discuss the case $\xco=r$ (as in (\ref{DimensionlessQuant})) and in particular explain how to express observables $\mathbb{O}[\con,\eps](\tco)$ when $\eps_r$ is only known as a function of $r$. From the undeformed Schwarzschild geometry we have the following explicit relations\footnote{We briefly comment on another possible example, namely $\xco=\rho$, the radial spatial distance to the horizon (which was used in some works of the EMD approach,  \emph{e.g.} \cite{DelPiano:2023fiw,DelPiano:2024nrl}). In this case we have
\begin{align}
&\rho|_{\con=0,\eps=0}(\tco)=\tfrac{a}{2}+\log\left(\tfrac{1}{2}\left(-2+(8+6\tco)^{2/3}+a\right)\right)\,,&&\text{with}&&a=\sqrt{(8+6\tco)^{2/3}(-4+(8+6\tco)^{2/3})}\,.
\end{align}
and $ \tco|_{\con=0,\eps_\xco=0}(\rho)$ requires inverting this relation: while (locally) this inverse exists, due to the transcendental nature of this equation, it can in general not be given in a simple closed form. We shall therefore not explicitly discuss this example in the following.
}
\begin{align}
&r|_{\con=0,\eps=0}(\tco)=\tfrac{1}{2}(8+6\tco)^{2/3}\,,&&\text{and}&&\tco|_{\con=0,\eps_r=0}(r)=\frac{1}{3}\left(\sqrt{2}\,r^{3/2}-4\right)\,,
\end{align}
which implies for the energy-density
\begin{align}
&\eps(\tco)\sim \eps_r\left(\tfrac{1}{2}(8+6\tco)^{2/3}\right)\,,&&\text{such that} &&\mathcal{I}_{n}[\eps](\tco)\sim\frac{d^{|n|}}{d\tco^{|n|}}\,\eps_r\left(\tfrac{1}{2}(8+6\tco)^{2/3}\right)\hspace{0.2cm}\forall n\leq 0\,.\label{rMomentaNeg}
\end{align}
Among the $\mathcal{I}_{n>0}[\eps](\tco)$ we shall concentrate on $n=1,2$: we can obtain leading order relations by substituting $\tcop\to \frac{\sqrt{2}(r')^{3/2}-4}{4}$ and changing integration variables 
\begin{align}
\int_0^{\tco}d\tco'\longrightarrow\int_{\rh}^{r(\tco)|_{\con=0,\eps=0}}\frac{dr'}{\sqrt{1-f}}\,,\label{IntegApprox}%\longrightarrow\int_{\rh}^r\sqrt{\frac{r'}{2}}\,dr'\,
\end{align}
such that we concretely obtain
\begin{align}
&\mathcal{I}_{1}[\eps](\tco)\sim \sqrt{2}\,\int_{\rh}^{\tfrac{1}{2}(8+6\tco)^{2/3}}\,\frac{r^{\prime 3/2}\,\eps_r(r')dr'}{\sqrt{1-f(r')}}\,,\nonumber\\
&\mathcal{I}_{2}[\eps](\tco)\sim\frac{\sqrt{2}}{(3\tco+4)}\,\int_{\rh}^{\tfrac{1}{2}(8+6\tco)^{2/3}}\frac{dr'}{\sqrt{1-f(r')}}\int_{\rh}^{r'}\,\frac{r^{\prime\prime 3/2}\,\eps_r(r'')dr''}{\sqrt{1-f(r'')}}\,.\label{rMomentaPos}
\end{align}
This prescription leads to exact integrals up to terms that are of order $\eps$ (and thus contribute terms to $\mathcal{I}_n[\eps]$ that go beyond the linear response), while at the same time not requiring to express $\tco$ as a function of $r$ (or its inverse) beyond the undeformed relation. We also remark that replacing $\rh\to 2$ in the integration boundaries or $\sqrt{1-f}\to \sqrt{\frac{2}{r}}$ in the integrands would also change the momenta by terms that are negligible in the linear response.

Equations (\ref{rMomentaNeg}) and (\ref{rMomentaPos}) allow to compute observables to linear order as functions of $\tco$ even in the case that the energy density is only given in terms of the radial coordinate. This scenario indeed occurs in numerous examples of spherically symmetric and static black holes in the literature (see the following Section~\ref{Sect:ExamplesMetrics}). In subsequent Sections, we shall further develop the linear response formalism to compute observables as functions of $\tco$ as in (\ref{LinearResponse}), with (\ref{LinearResponseXmod}) and (\ref{LinearResponseInverseXmod}) allowing to reparametrise it in terms of other quantities. However, due to the common use of the variable $r$ in many works in the literature, we shall provide certain observables also in the $r$-parametrisation.

%%%%%%%%%%%%%%%%%%%%

\subsection{Examples of Deformed Black Hole Geometries}\label{Sect:ExamplesMetrics}
In order to illustrate the result (\ref{LinearResponse}) for concrete observables in subsequent Sections, we

\begin{wrapfigure}{r}{0.45\textwidth}
\vspace{-0.75cm}
\begin{center}
\includegraphics[width=7.5cm]{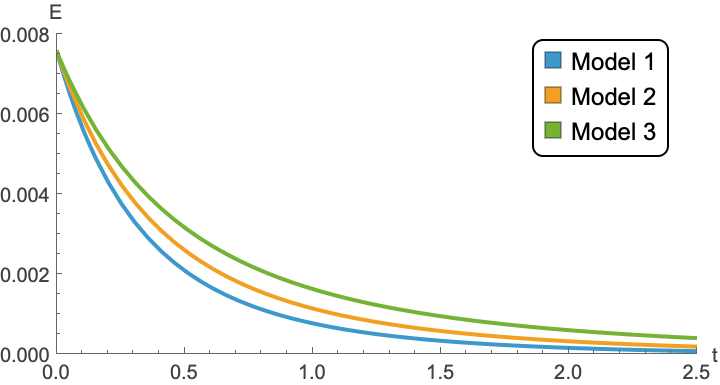}
\end{center}
\vspace{-0.5cm}
\caption{Energy density $\eps$ as a function of $\tco$ for the black holes in table~\ref{Tab:OverviewModels}. The deformation parameters $\gamma$ have been chosen such that $\eps(\tco=0)$ agrees for all three black holes, \emph{i.e.} concretely $\gamma=(0.01,0.015,0.029)$ for the three models.}
\label{Fig:CompEpsilonDirect}
\vspace{-0.2cm}
\end{wrapfigure}

%\begin{figure}[h]
%\begin{center}
%\includegraphics[width=10cm]{EpsModels.png}
%\end{center}
%\caption{Energy density $\eps$ as a function of $\tco$ for the black holes in table~\ref{Tab:OverviewModels}. The deformation parameters $\gamma$ have been chosen such that $\eps(\tco=0)$ agrees for all four black holes, \emph{i.e.} concretely $\gamma=(0.01,0.015,0.029)$ for the Hayward-, Simpson-Visser I- and Simpson-Visser II-black hole.}
%\label{Fig:CompEpsilonDirect}
%\end{figure}
\noindent
shall consider a number examples for $\eps$. % that can be related to metric functions which have appeared in the literature. 
We have summarised these functions in Table~\ref{Tab:OverviewModels}. These examples are constructed in a way to reproduce metric-deformation functions that have appeared in certain models of deformed black hole geometries.\footnote{Model 1 is the metric function of the black hole geometry introduced by Hayward in \cite{Hayward:2005gi}, while Model 2 yields the metric function for a geometry proposed by Simpson and Visser in \cite{Simpson:2018tsi}. Finally, model 3 is inspired by a metric function appearing in \cite{Simpson:2019mud}: we remark, however, that the full geometry considered there is slightly more complicated. All 3 models have also been studied in \cite{DelPiano:2024nrl}.} For better visualisation, the energy density $\eps$ is plotted (as a function of $\tco$) for all three models in Figure~\ref{Fig:CompEpsilonDirect}, where the value of $\gamma$ has been chosen in such a way that the value at the horizon agrees for all three cases. As shown in the second column of Table~\ref{Tab:OverviewModels}, due to their connection to geometries already studied in the literature, for all models $\eps$ is naturally expressed in  terms of the (dimensionless) radial variable $r$ (called $\eps_r$). While in all three cases $\tco$ can analytically be calculated as a function of $r$,\footnote{We remark, however, that the inversion of this results (\emph{i.e.} expressing $r$ as a function of $\tco$) is more complicated.} we refrain from providing the explicit expressions, as they will not be relevant in the following. In the fourth column of Table~\ref{Tab:OverviewModels} we have provided the leading order of $\eps$ in $\gamma$ as a function of $\tco$ to highlight a relative similarity of all three models that is also evident from Figure~\ref{Fig:CompEpsilonDirect}: indeed, the energy density of all models can be very well approximated to leading order by functions of the form
\begin{align}
&\eps(\tco)=\frac{\alpha}{(8+6\tco)^\beta}\sim \frac{\alpha}{(2r)^{3\beta/2}}\,,&&\text{with} &&\alpha,\beta\in\mathbb{R}_+\,,\label{PowerFunctionEps}
\end{align}
for $\alpha\ll 1$ such that concretely $\sim$ indicates corrections of order $\mathfrak{o}(\alpha)$. %We shall use this form for certain computations, in order to capture conceptual trends of a wider class of examples.

\begin{figure}[h!]
\begin{center}
\includegraphics[width=7.5cm]{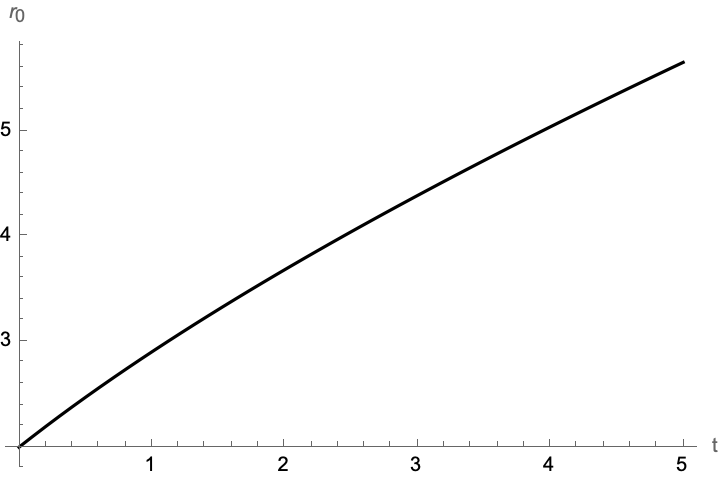}\hspace{1cm}\includegraphics[width=7.5cm]{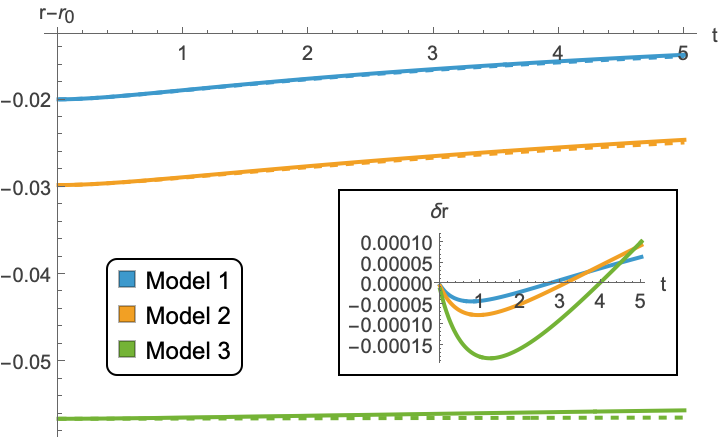}
\end{center}
\caption{Left panel: the radial coordinate $r_0=r(\tco)\big|_{\con=0,\eps=0}=\tfrac{1}{2}(8+6\tco)^{2/3}$ as a function of $\tco$ in the undeformed case. Right panel: linear corrections $ r(\tco)-r(\tco)\big|_{\con=0,\eps=0}$ for the three models introduced in this Section (and for the same values of $\gamma$ as in Figure~\ref{Fig:CompEpsilonDirect}). For each model, we have plotted the exact correction to $r(\tco)$ (dashed line) calculated numerically, the linear response (\ref{LinRespPos}) with the exact momenta $\mathcal{I}_n[\eps]$ calculated numerically (heavy solid line) and the linear response (\ref{LinRespPos}) with the momenta $\mathcal{I}_n[\eps]$ as in Section~\ref{App:MomentaExamples} (thin solid line). For better visibility, the inset graphic provides the numerical difference between the latter two. }
\label{Fig:Position_Examples}
\end{figure}

\begin{figure}[h]
\begin{center}
\includegraphics[width=7.5cm]{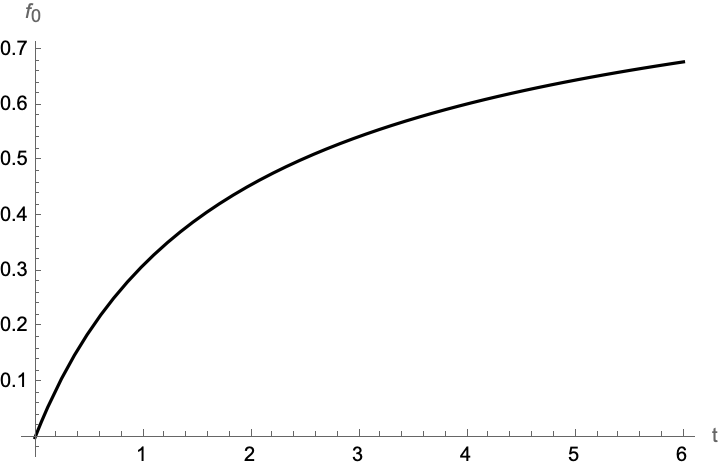}\hspace{1cm}\includegraphics[width=7.5cm]{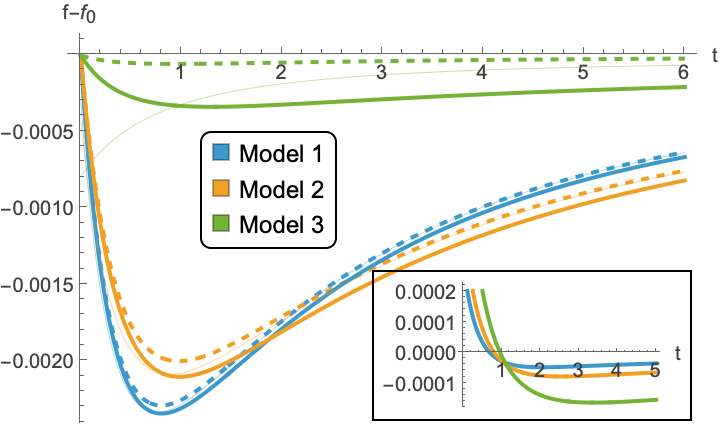}
\end{center}
\caption{Left panel: metric function $f_0=f(\tco)\big|_{\con=0,\eps=0}=1-\tfrac{4}{(8+6\tco)^{2/3}}$ as a function of $\tco$ in the undeformed case. Right panel: linear corrections $ f(\tco)-f(\tco)\big|_{\con=0,\eps=0}$ for the three models introduced in this Section (and for the same values of $\gamma$ as in Figure~\ref{Fig:CompEpsilonDirect}). For each model, we have plotted the exact correction to $f(\tco)$ (dashed line) calculated numerically, the linear response (\ref{LinRespPos}) with the exact momenta $\mathcal{I}_n[\eps]$ calculated numerically (heavy solid line) and the linear response (\ref{LinRespPos}) with the momenta $\mathcal{I}_n[\eps]$ as in Section~\ref{App:MomentaExamples} (thin solid line). For better visibility, the inset graphic provides the numerical difference between the latter two. }
\label{Fig:Fmet_Examples}
\end{figure}

Given $\eps$ for the three models, we can calculate the relevant momenta for the examples in Table~\ref{Tab:OverviewModels}, which for convenience are summarised in Appendix~\ref{App:MomentaExamples}. These momenta in turn can be used to directly calculate linear responses for physically relevant invariants as in (\ref{LinearResponse}). For example, Figures~\ref{Fig:Position_Examples} and \ref{Fig:Fmet_Examples} compare the exact, numerical form of the radial coordinate $r$ and the metric function $f$ to the linear responses of the form (\ref{LinearResponse}). As is evident, at least for small values of $\gamma$ (such that $\eps$ and its momenta are small throughout space-time), the linear response is an excellent approximation to the exact numerical result.

\begin{table}[t]
\begin{center}
\begin{tabular}{|c||c|c|c|c|}\hline
&&&&\\[-10pt]
{\bf Model} & $\eps_r(r)$ & $f(r)=h(r)$ &  leading order $\eps(\tco)$ & Reference\\[2pt]\hline\hline
&&&&\\[-10pt]
1 & $\frac{48\gamma(1+\gamma)^3}{(8\gamma+r^3(1+\gamma)^3)^2}$ & $1-\frac{2r^2}{r^3+8\frac{\gamma}{(1+\gamma)^3}}$   & $\frac{3072\,\gamma}{(8+6\tco)^4}+\mathcal{O}(\gamma^2)$ & \cite{Hayward:2005gi}\\[12pt]
2 &  $\frac{8\gamma(1+\gamma)(2+\gamma)}{r^2\left(r^2(1+\gamma)^2+4\gamma(2+\gamma)\right)^{3/2}}$ & $1-\frac{2}{\sqrt{r^2+4\frac{\gamma(2+\gamma)}{(1+\gamma)^2}}}$   & $\frac{512\,\gamma}{(8+6\tco)^{10/3}}+\mathcal{O}(\gamma^2)$ &  \cite{Simpson:2018tsi} \\[12pt]
3 & $\frac{4\log(1+\gamma)}{r^4(1+\gamma)^{1+\frac{2}{r(1+\gamma)}}}\,$ & $1-\frac{2}{r}\,\text{exp}\left(-\tfrac{2}{r}\frac{\log(1+\gamma)}{1+\gamma}  \right)$    & $\frac{64\,\gamma}{(8+6\tco)^{8/3}}+\mathcal{O}(\gamma^2)$ & \cite{Simpson:2019mud}\\[12pt]\hline
\end{tabular}
\end{center}
\caption{Overview over three black hole models which shall be discussed throughout this work: the second column shows $\eps$ as a function of the radial variable $r$. Indeed, all examples are chosen in such a way that the reproduce metric functions (third column), which have appeared in the literature (with the reference indicated in the last column). Assuming $\gamma$ to be a small parameter, the fourth column gives the leading term in $\gamma$ of $\eps$ as a function of $\tco$, which follows a power law as in (\ref{PowerFunctionEps}). The convention for $\gamma$ is such that the zeroes of $f$ (\emph{i.e.} the position of the horizon) is located at $\rh=\frac{2}{1+\gamma}$. }
\label{Tab:OverviewModels}
\end{table}

%%%%%%%%%%%%%%%%%%%%%%%%%%%%%%%%%%%%%%%
\section{Structure of the Geometry and Geodesic Motion}\label{Sect:GeodesicMotionSect}
In this Section, we first study (curvature) invariants that are related to the structure of the space-time itself: we show how to re-write the Ricci- and Kretschmann-scalar in the form (\ref{LinearResponse}) for a generic point outside the event horizon. In the same manner, we provide the linear response expression for the Hawking temperature, which we define as the surface gravity on the event horizon, from which we can compute the entropy of the black hole. In a second step (Section~\ref{Sect:GeodesicMotion}), we consider geodesic motions of particles around the black hole: we particularly present expressions of the form (\ref{LinearResponse}) for characteristic quantities of spherical orbits of massless- and massive particles, such as the radius of the photon sphere and the black hole shadow, as well as the position of the innermost stable circular orbit (ISCO), respectively.
%%%%%%%%%
\subsection{Curvature Invariants and Hawking Temperature}\label{Sect:Hawking}
\begin{figure}[h]
\begin{center}
\includegraphics[width=7.5cm]{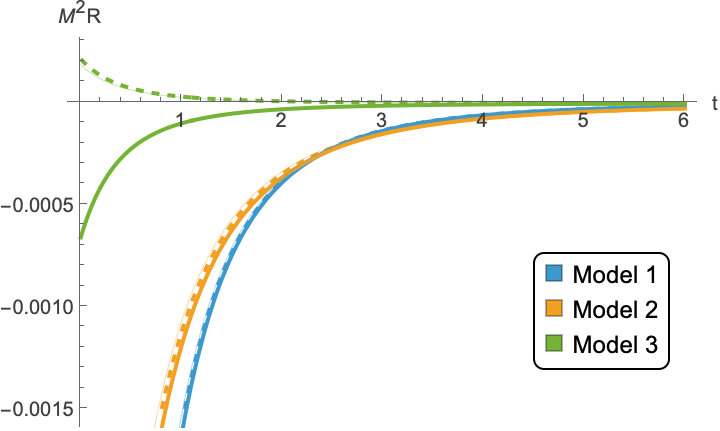}\hspace{1cm}\includegraphics[width=7.5cm]{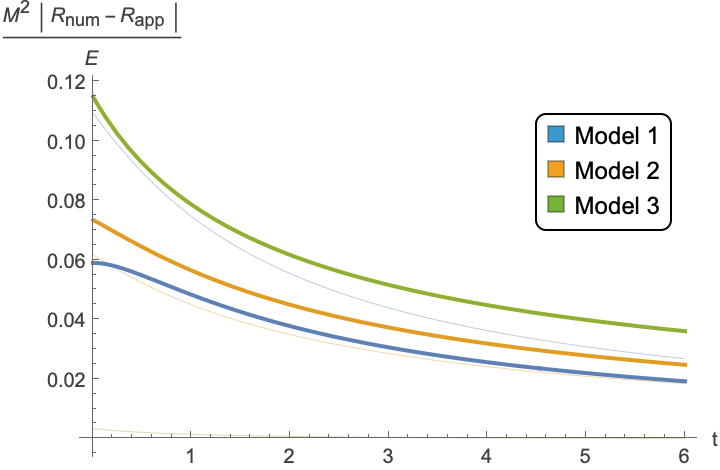}
\end{center}
\caption{Left panel: Ricci scalar for three examples introduced in Section~\ref{Sect:ExamplesMetrics} (for the same values of $\gamma$ as in Figure~\ref{Fig:CompEpsilonDirect}).  For each model, we have plotted the exact correction to $r(\tco)$ (dashed line) calculated numerically, the linear response (\ref{LinRespPos}) with the exact momenta $\mathcal{I}_n[\eps]$ calculated numerically (heavy solid line) and the linear response (\ref{LinRespPos}) with the momenta $\mathcal{I}_n[\eps]$ as in Section~\ref{App:MomentaExamples} (thin solid line). Right panel: in order to estimate the error of the approximation, we have plotted for the linear responses (thin and heavy solid line as in the left panel) the difference to the (exact) numerical value in units of $\eps$. As is evident in all cases, the error made by the linear approximation is of the order of a few percent of the energy density.}
\label{Fig:Ricci_Examples}
\end{figure}

\begin{figure}[h]
\begin{center}
\includegraphics[width=7.5cm]{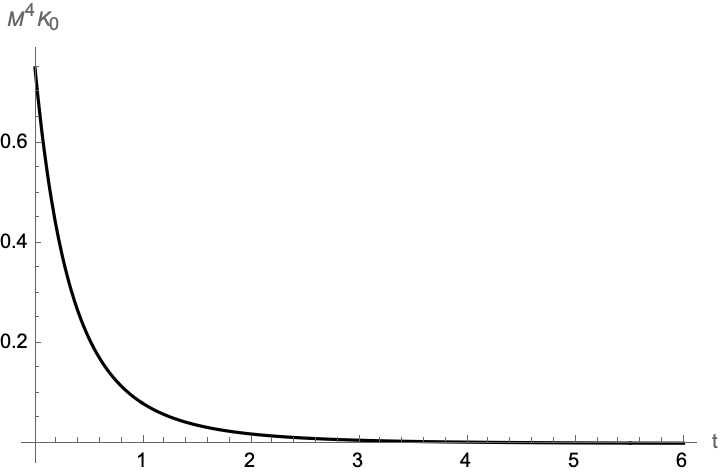}\hspace{1cm}\includegraphics[width=7.5cm]{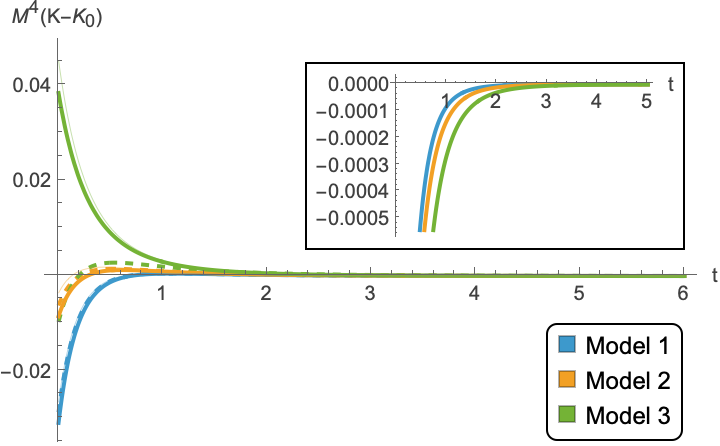}
\end{center}
\caption{Left panel: Kretschmann scalar $M^4\,K_0=M^4\,K(\tco)\big|_{\con=0,\eps=0}=\tfrac{192}{(4+3\tco)^{4}}$ as a function of $\tco$ in the undeformed case. Right panel: linear corrections $ M^4\,(K(\tco)-K(\tco)\big|_{\con=0,\eps=0})$ for the three models introduced in Section~\ref{Sect:ExamplesMetrics} (and for the same values of $\gamma$ as in Figure~\ref{Fig:CompEpsilonDirect}). For each model, we have plotted the exact correction to $f(\tco)$ (dashed line) calculated numerically, the linear response (\ref{LinRespPos}) with the exact momenta $\mathcal{I}_n[\eps]$ calculated numerically (heavy solid line) and the linear response (\ref{LinRespPos}) with the momenta $\mathcal{I}_n[\eps]$ as in Section~\ref{App:MomentaExamples} (thin solid line). For better visibility, the inset graphic provides the numerical difference between the latter two. }
\label{Fig:Kretschmann_Examples}
\end{figure}

Using the form of the metric function (\ref{FormMetric}), the Ricci-scalar $R$ and Kretschmann-scalar $K$ can be written (in dimensionless form) as
\begin{align}
&M^2R=\frac{2-2f-4r\,\frac{df}{dr}-r^2 \frac{d^2f}{dr^2}}{r^2}\,,&&\text{and}&&M^4K=\frac{4-8f-4f^2+4r^2\left(\frac{df}{dr}\right)^2+r^4\left(\frac{d^2f}{dr^2}\right)^2}{r^4}\,.
\end{align}
We can write the linear response for both invariants in the form (\ref{FormFcomp}), \emph{i.e.} 
\begin{align}
&M^2\,R(\tco)\sim \sum_{n=-1}^0\coff{R}{n}(\tco)\,\mathcal{I}_n[\eps](\tco)\,,&&\text{with} &&\left\{\begin{array}{l} \coff{R}{0}(\tco)=4\,,\\[4pt]\coff{R}{-1}(\tco)=\frac{4+3\tco}{2}\,, \end{array}\right.\label{LinRicci}
\end{align}
and 
\begin{align}
&M^4\,K(\tco)\sim M^4\,K(\tco)\big|_{\con=0,\eps=0}+\coff{K}{\con}(\tco)\,\con+\sum_{n=-1}^2\coff{K}{n}(\tco)\,\mathcal{I}_n[\eps](\tco)\,,\label{LinKretsch}
\end{align}
with 
\begin{align}
&M^4\,K(\tco)\big|_{\con=0,\eps=0}=\coff{K}{1}(\tco)=\frac{3\cdot 2^{10}}{\sh^4}\,,&&\coff{K}{\con}(\tco)=-\frac{3\cdot 2^{14}}{\sh^5}\,,&&\coff{K}{-1}(\tco)=\frac{2^4}{\sh}\,,\nonumber\\
&\coff{K}{0}(\tco)=-\frac{2^7}{\sh^2}\,,&&\coff{K}{2}(\tco)=-\frac{3^2\cdot 2^{10}}{\sh^4}\,.
\end{align}
where for convenience we have introduced the shorthand notation
\begin{align}
\sh=8+6\tco\,.\label{ShorthandQnotation}
\end{align}
%\begin{align}
%&M^4\,K(\tco)\big|_{\con=0,\eps=0}=\frac{192}{(4+3\tco)^4}\,,&&\coff{K}{\con}(\tco)=-\frac{1536}{(4+3\tco)^5}\,,&&\coff{K}{-1}(\tco)=\frac{8}{4+3\tco}\,,\nonumber\\
%&\coff{K}{0}(\tco)=-\frac{2^7}{(4+3\tco)^2}\,,&&\coff{K}{1}(\tco)=\frac{192}{(4+3\tco)^4}\,,&&\coff{K}{2}(\tco)=-\frac{576}{(4+3\tco)^4}\,.
%\end{align}
We first remark that $\coff{K}{\con}(\tco=0)=-\tfrac{3}{2}\neq 0$. Notice that this allows to relate to the constant $\con$ to the value of the Kretschmann scalar at the horizon, \emph{i.e.}
\begin{align}
\con\sim\frac{1}{6}\left(3-4\,M^4\,K(0)-8\,\mathcal{I}_0[\eps](0)+8\,\mathcal{I}_{-1}[\eps](0)\right)\,.\label{rhKretschRel}
\end{align} 
The results (\ref{LinRicci}) and (\ref{LinKretsch}) are showcased for the examples introduced in Section~\ref{Sect:ExamplesMetrics}, in Figure~\ref{Fig:Ricci_Examples} and \ref{Fig:Kretschmann_Examples} respectively. In the former case, we remark that the Ricci scalar vanishes identically in the undeformed case, such that the linear response constitutes the leading contribution.

\noindent
As is evident in particular for model 3 (green curve in  Figure~\ref{Fig:Ricci_Examples}), even though the error is of the order of a few percent, such observables can still be very susceptible to higher order corrections. These are evidently less relevant in the case where already the undeformed observable is non-trivial (as in the case of the Kretschmann scalar in Figure~\ref{Fig:Kretschmann_Examples}).

A further invariant that is related to the geometry of the black hole at the horizon is the 

\begin{wrapfigure}{r}{0.45\textwidth}
\vspace{-0.55cm}
\begin{center}
\includegraphics[width=7.5cm]{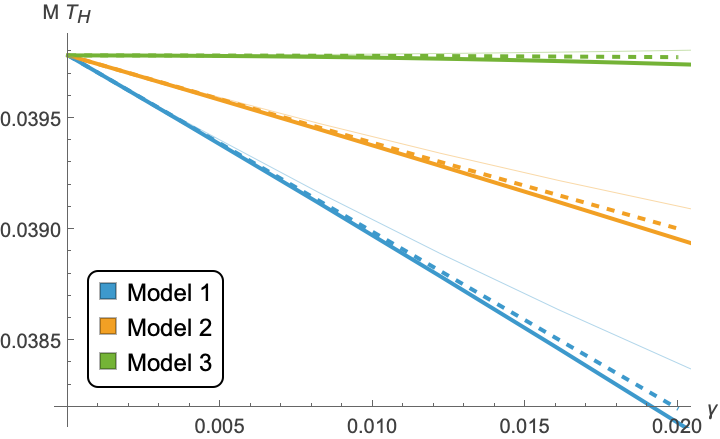}
\end{center}
\vspace{-0.5cm}
\caption{Hawking Temperature $T_H$ as a function of $\gamma$ for the three models introduced in Section~\ref{Sect:ExamplesMetrics}. We have numerically calculated the exact expression (dashed line), the linear response (\ref{LinRespPos}) with the exact momenta $\mathcal{I}_n[\eps]$ calculated numerically (heavy solid line) and the linear response (\ref{LinRespPos}) with the momenta $\mathcal{I}_n[\eps]$ as in Section~\ref{App:MomentaExamples} (thin solid line).  }
\label{Fig:Temp_Examples}
\vspace{-1cm}
\end{wrapfigure}

\noindent
Hawking temperature, which is defined in terms of the surface gravity $\kappa=\tfrac{1}{2M}\,\tfrac{df}{dr}|_{r=\rh}$ as follows
\begin{align}
&M\,T_H=\frac{M\kappa}{2\pi}\sim \frac{1}{8\pi}+\coff{T_H}{\con}\,\con+\coff{T_H}{0}\,\mathcal{I}_0[\eps](0)\,,\label{HawkingTemp}
\end{align}
with
\begin{align}
&\coff{T_H}{\con} = -\frac{1}{16\pi}\,, &&\text{and}&& \coff{T_H}{0}=-\frac{1}{2\pi}\,.
\end{align}
Notice, that for this invariant, the coefficients $\coff{T_H}{\con}$ and $\coff{T_H}{0}$ are in fact constants (which is why we have dropped the argument $\tco$), while $\mathcal{I}_0[\eps](0)=\eps(0)$ is evaluated at the horizon. Similar to (\ref{rhKretschRel}), the relation (\ref{HawkingTemp}) provides another way of expressing the constant $\con$ in terms of a physical quantity, namely the temperature of the black hole. 

Furthermore, the temperature (\ref{HawkingTemp}) allows to compute the linear response to the entropy of the black hole
\begin{align}
S&=\int\frac{dM}{T_H}\sim 4\pi\,M^2\,+2\pi\,M^2\,\con+32\pi\,\int\,M\,\eps(0)\,dM\,,\nonumber
\end{align}
where $\eps(0)$ may implicitly depend on the mass of the black hole.

Finally, we remark that the temperature (\ref{HawkingTemp}) can be computed \emph{exactly}, using for example the EMD approach \cite{Binetti:2022xdi,DelPiano:2023fiw,DelPiano:2024nrl,SH}
\begin{align}
M\,T_H=\frac{1}{4\pi \rh}-\frac{\rh}{4\pi}\,\eps(0)\,.
\end{align}
From this relation it becomes evident that the approximation in (\ref{HawkingTemp}) stems entirely from expanding the position of the event horizon $\rh= 2+\con$, which is in fact independent of any chosen parametrisation.

\subsection{Geodesic Motion}\label{Sect:GeodesicMotion}
A brief overview of geodesic movement in spherically symmetric geometries is given in Appendix~\ref{App:GeodesicMotion}. In the following we shall mostly be interested in circular orbits around the black hole by massless (\emph{i.e.} photons) and massive particles. 

\subsubsection{Circular Orbits of Massless Particles: Shadow and Lyapunov Exponent}
\begin{figure}[h]
\begin{center}
\includegraphics[width=7.5cm]{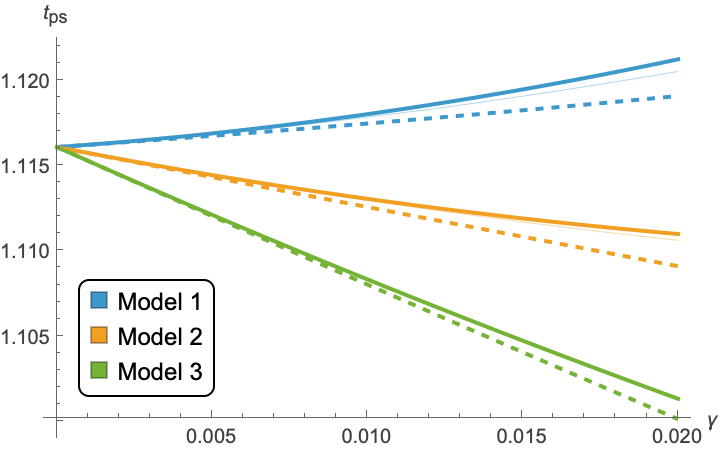}\hspace{1cm}\includegraphics[width=7.5cm]{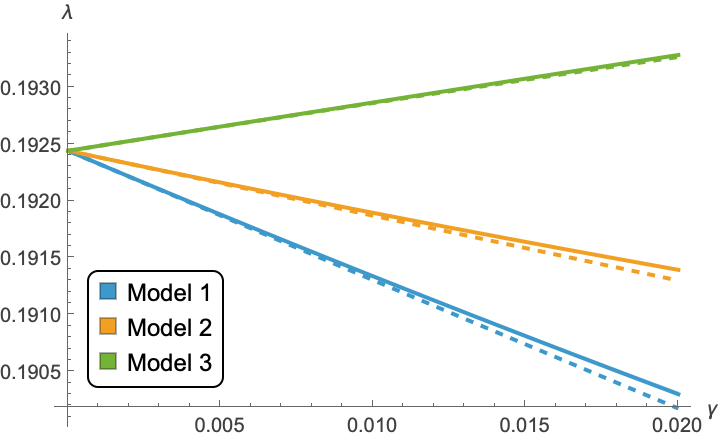}
\end{center}
\caption{Position of the photon sphere $\tco_{\text{ps}}$ (left panel) and Lyapunov exponent $\lambda$ (right panel) as a function of $\gamma$ for the three models introduced in Section~\ref{Sect:ExamplesMetrics}. We have numerically calculated the exact expression (dashed line), the linear response (\ref{LinRespPos}) with the exact momenta $\mathcal{I}_n[\eps]$ calculated numerically (heavy solid line) and the linear response (\ref{LinRespPos}) with the momenta $\mathcal{I}_n[\eps]$ as in Section~\ref{App:MomentaExamples} (thin solid line).  }
\label{Fig:Photon_Examples}
\end{figure}

Following \cite{Rezzolla2016,DelPiano:2024nrl} (see also eq.~(\ref{PotentialRadial}) in Appendix~\ref{App:GeodesicMotion}), we can introduce an effective potential to describe the geodesic motion of massless particles
\begin{align}
V_{\text{massless}}(\tco)=\frac{f}{r^2}\sim V_{\text{massless}}(\tco)\big|_{\con=0,\eps=0}+\coff{V_{\text{massless}}}{\con}(\tco)\,\con+\sum_{n=1}^2\coff{V_{\text{massless}}}{n}(\tco)\,\mathcal{I}_n[\eps](\tco)\,,
\end{align}
with 
\begin{align}
&V_{\text{massless}}(\tco)\big|_{\con=0,\eps=0}=\frac{4}{\sh^2}(\sh^{2/3}-4)\,,&&\coff{V_{\text{massless}}}{\con}(\tco)=\frac{384-4\sh^{2/3}(16+\sh)}{3\sh^3}\,,&&\coff{V_{\text{massless}}}{1}(\tco)=-\frac{8}{\sh^2}\,,\nonumber\\
&\coff{V_{\text{massless}}}{1}(\tco)=\frac{4(6-\sh^{2/3})}{\sh^2}\,.
\end{align}
Extrema of this potential represent circular orbits: as explained in Appendix~\ref{App:GeodesicMotion}, in the undeformed case, the Schwarzschild geometry has a single maximum, which is the so-called photon sphere. In the presence of the deformation $\eps$, to linear order the position of this maximum is
\begin{align}
&\tco_{\text{ps}}\sim\tco_{\text{ps}}\big|_{\con=0,\eps=0}+ \coff{\tco_{\text{ps}}}{\con}\,\con+\sum_{n=1}^2\coff{\tco_{\text{ps}}}{n}\,\mathcal{I}_n[\eps]\left(\tco_{\text{ps},0}\right)\,,&&\text{with}&&\left\{\begin{array}{l}\tco_{\text{ps}}\big|_{\con=0,\eps=0}=\tco_{\text{ps},0}:=\frac{3\sqrt{6}-4}{3}\,,\\[4pt]\coff{\tco_{\text{ps}}}{\con}=\frac{3\sqrt{6}-4}{6}\,,\\[4pt]\coff{\tco_{\text{ps}}}{0}=-\frac{27\sqrt{3}}{2^{3/2}}\,,\\[4pt]\coff{\tco_{\text{ps}}}{1}=\frac{3\sqrt{3}}{2^{3/2}}\,,\\[4pt]\coff{\tco_{\text{ps}}}{2}=-\frac{3\sqrt{3}}{2^{3/2}}\,.\label{PositionPhotonSphere}
\end{array}\right.
\end{align}
The position for the photon sphere $\tco_{\text{ps}}$ in the case of three different models is depicted in the left panel of Figure~\ref{Fig:Photon_Examples}. The radius of the black hole shadow \cite{Perlick:2021aok,DelPiano:2024nrl}, as observed from a distant observer can be computed as
\begin{align}
\frac{b_{\text{sh}}}{M}=(V_{\text{massless}}(\tco_{\text{ps}}))^{-1/2}\sim 3\sqrt{3}+ \coff{b_{\text{ps}}}{\con}\,\con+ \coff{b_{\text{ps}}}{1}\,\mathcal{I}_1[\eps]\left(\tco_{\text{ps},0}\right)\,,&&\text{with} &&\coff{b_{\text{ps}}}{\con}=\coff{b_{\text{ps}}}{1}=\frac{3\sqrt{3}}{2}\,.\label{BHshadow}
\end{align}
This observable has already been studied in the linear approach in \cite{DelPiano:2024nrl}, notably for the three models introduced in Section~\ref{Sect:ExamplesMetrics}. Here we remark on two more invariants that shall be important later on, namely 
\begin{align}
M\,\omega_c=(V_{\text{massless}}(\tco_{\text{ps}}))^{1/2}\sim\frac{1}{3\sqrt{3}}+ \coff{\omega_c}{\con}\,\con+ \coff{\omega_c}{1}\,\mathcal{I}_1[\eps]\left(\tco_{\text{ps},0}\right)\,,&&\text{with} &&\coff{\omega_c}{\con}=\coff{\omega_c}{1}=-\frac{1}{6\sqrt{3}}\,,\label{Mogc}
\end{align}
and the \emph{Lyapunov exponent} \cite{Cardoso:2008bp}
\begin{align}
&\lambda=\sqrt{-\frac{1}{2}\frac{r^2}{f}\left(\frac{d^2}{dr_*^2}\frac{f}{r^2}\right)}\bigg|_{r=r_{\text{ps}}}\,,&&\text{with} &&\frac{dr}{dr_{*}}=f\,.
\end{align}
In the linear response approach, we obtain
\begin{align}
\lambda\sim\frac{1}{3\sqrt{3}}+ \coff{\lambda}{\con}\,\con+ \sum_{n=\pm 1}\coff{\omega_c}{n}\,\mathcal{I}_n[\eps]\left(\tco_{\text{ps},0}\right)\,,&&\text{with} &&\left\{\begin{array}{l}\coff{\lambda}{\con}=\coff{\lambda}{1}=-\frac{1}{6\sqrt{3}}\,, \\ \coff{\lambda}{-1}=\frac{9}{4\sqrt{2}}\,.\end{array}\right.\label{Lyapunov}
\end{align}

\noindent
The Lyapunov exponent is schematically plotted in the right panel of Figure~\ref{Fig:Photon_Examples} for various examples. The quantities $\omega_c$ and $\lambda$ can be related to the eikonal limit of scalar quasi-normal modes \cite{Cardoso:2008bp}, as we shall discuss in Section~\ref{Sect:QNM}. Before turning to invariants related to geodesic motions of massive particles, we remark the following form of the \emph{critical exponent} $\gamma_c$, which was first introduced in \cite{Pretorius:2007jn}, in the linear response formalism
\begin{align}
&\gamma_c:=\frac{M\omega_c}{2\pi\lambda}\sim\frac{1}{2\pi}+\coff{\gamma_c}{-1}\,\mathcal{I}_{-1}[\eps]\left(\tfrac{3\sqrt{6}-4}{3}\right)\,,&&\text{with} &&\coff{\gamma_c}{-1}=-\frac{27\sqrt{3}}{8\pi\sqrt{2}}\,.
\end{align}

%%%%%%%%

\subsubsection{Circular Orbits of Massive Particles}
\begin{figure}[h]
\begin{center}
\includegraphics[width=7.5cm]{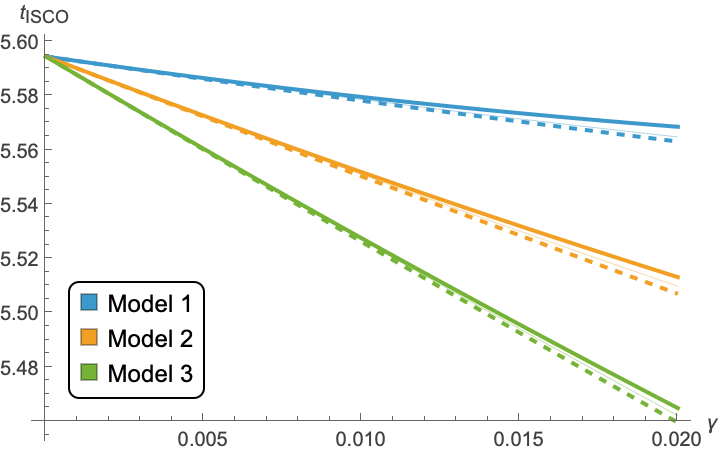}\hspace{1cm}\includegraphics[width=7.5cm]{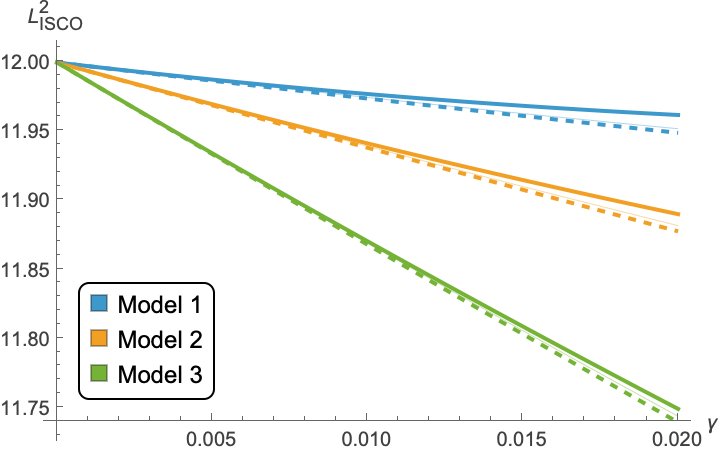}
\end{center}
\caption{Position of the ISCO $\tco_{\text{ISCO}}$ (\ref{Tisco}) left panel) and (square of the) critical value of the  angular momentum $L_{\text{ISCO}}^2$ (\ref{CriticalAngular}) (right panel) as a function of $\gamma$ for the three models introduced in Section~\ref{Sect:ExamplesMetrics}. We have numerically calculated the exact expression (dashed line), the linear response (\ref{LinRespPos}) with the exact momenta $\mathcal{I}_n[\eps]$ calculated numerically (heavy solid line) and the linear response (\ref{LinRespPos}) with the momenta $\mathcal{I}_n[\eps]$ as in Section~\ref{App:MomentaExamples} (thin solid line).  }
\label{Fig:ISCO_Examples}
\end{figure}

The geodesic motion of massive particles is described using the following effective potential (see \emph{e.g.} \cite{Rezzolla2016,Jha:2024ltc} as well as Appendix~\ref{App:GeodesicMotion})
\begin{align}
V_{\text{mass}}=f\,\left(1+\frac{L^2}{r^2}\right)\,,\label{PotIsco}
\end{align}
where $L$ is the angular momentum of the test particle (in units of $M$) that orbits the black hole. To linear order, we have
\begin{align}
V_{\text{mass}}(\tco)\sim\frac{(\sh^{2/3}-4)(4L^2+\sh^{4/3})}{\sh^2}+ \coff{V_{\text{mass}}}{\con}(\tco)\,\con+ \sum_{n= 1}^2\coff{V_{\text{mass}}}{n}(\tco)\,\mathcal{I}_n[\eps]\left(\tco\right)\,,
\end{align}
with
\begin{align}
&\coff{V_{\text{mass}}}{\con}(\tco)=\frac{4((8-\sh)\sh^{4/3}+L^2(96-\sh^{2/3}(16+\sh)))}{3\sh^3}\,,&&\coff{V_{\text{mass}}}{1}(\tco)=-\frac{2(4L^2+\sh^{4/3})}{\sh^2}\,,\nonumber\\
&\coff{V_{\text{mass}}}{2}(\tco)=-\frac{2(2L^2(6-\sh^{2/3})+\sh^{4/3})}{\sh^2}\,.
\end{align}
Circular orbits correspond to extrema of $V_{\text{mass}}$ and may be stable or unstable: here we shall be interested in the \emph{innermost stable circular orbit} (ISCO for short), which corresponds to a saddle point of the potential (see eq.~(\ref{IscoCondition})). Concretely, in the linear response, the ISCO is located at position $\tco_{\text{ISCO}}$ 
\begin{align}
&\tco_{\text{ISCO}}\sim\frac{4(3\sqrt{3}-1)}{3}+\coff{\tco_{\text{ISCO}}}{\con}\,\con+ \sum_{n=-1}^2\coff{\tco_{\text{ISCO}}}{n}\,\mathcal{I}_n[\eps]\left(\tfrac{4(3\sqrt{3}-1)}{3}\right)\,,&&\text{with} &&\left\{\begin{array}{l}\coff{\tco_{\text{ISCO}}}{\con}=\frac{2(3\sqrt{3})-1}{3}\,, \\[4pt] \coff{\tco_{\text{ISCO}}}{-1}=6^5\,, \\[4pt] \coff{\tco_{\text{ISCO}}}{0}=2\cdot 6^3\sqrt{3}\,, \\[4pt] \coff{\tco_{\text{ISCO}}}{1}=3\sqrt{3}\,, \\[4pt] \coff{\tco_{\text{ISCO}}}{2}=-3\sqrt{3}\,.\end{array}\right.\label{Tisco}
\end{align}
for a particle with the critical angular momentum $L^2_{\text{ISCO}}$
\begin{align}
&L^2_{\text{ISCO}}\sim 12+\coff{L^2_{\text{ISCO}}}{\con}\,\con+ \sum_{n=0}^1\coff{L^2_{\text{ISCO}}}{n}\,\mathcal{I}_n[\eps]\left(\tfrac{4(3\sqrt{3}-1)}{3}\right)\,,&&\text{with} &&\left\{\begin{array}{l}\coff{L^2_{\text{ISCO}}}{\con}=12\,,\\ [4pt]\coff{L^2_{\text{ISCO}}}{0}=-12^3\,, \\[4pt] \coff{L^2_{\text{ISCO}}}{1}=12\,.\end{array}\right.\label{CriticalAngular}
\end{align} 
 These results are plotted in Figure~\ref{Fig:ISCO_Examples} for three examples of black holes.

%%%%%%%%%%%%%%%%%%%%%%%%%%%%%%%%%%%%%%%%%%%%%%
\section{Scalar Field Dynamics and Quasi-normal Modes}\label{Sect:QNM}
As another example of local physics in the vicinity of the deformed black hole, we consider the dynamics of test fields. Concretely, we study the spectrum of quasi-normal modes of massless scalar fields in the background of the black hole geometry.
\subsection{Potential}
\begin{figure}[h]
\begin{center}
\includegraphics[width=5.5cm]{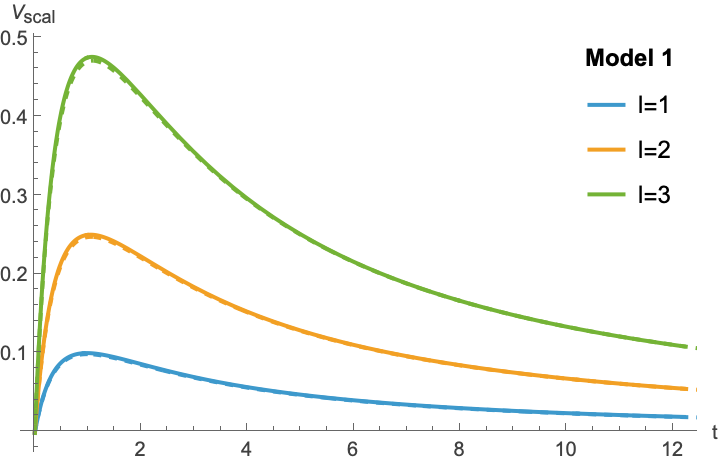}\hspace{0.25cm}\includegraphics[width=5.5cm]{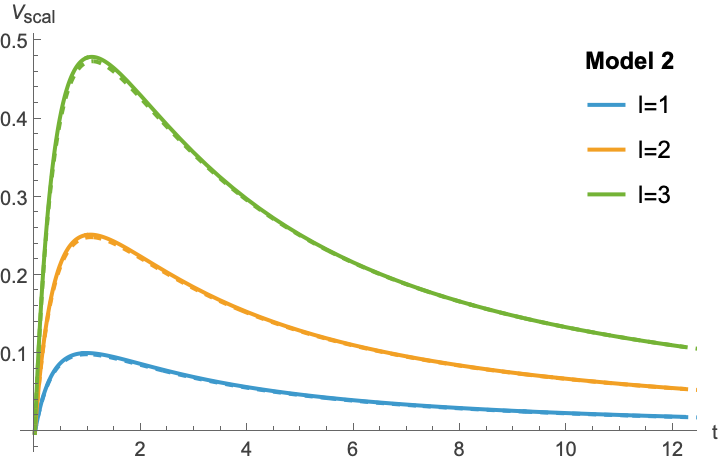}\hspace{0.25cm}\includegraphics[width=5.5cm]{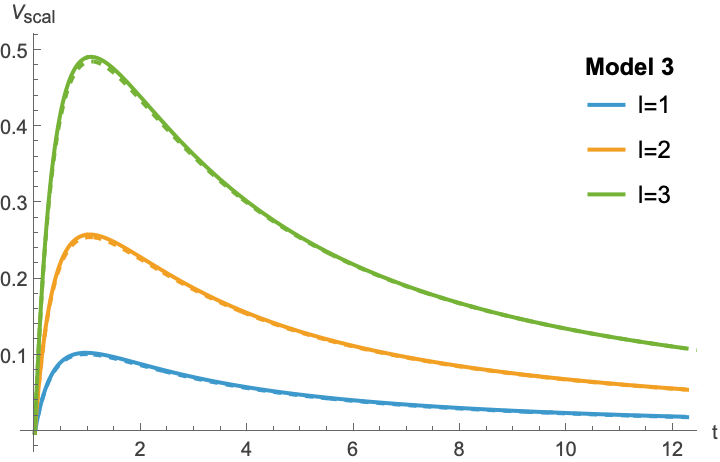}
\end{center}
\caption{Plot of the potential $V_\scal$ for the Hayward (left panel), the Simpson-Visser I (middle panel) and Simpson-Visser II (right panel) black hole, for different values of $\ell$: the solid lines represent the analytic (linear) approximations (\ref{VscalLin}) (using the momenta given in (\ref{HaywardMomenta}), (\ref{SV1Momenta}) and (\ref{SV2Momenta}) respectively), while the dashed lines represent numerical computations for the exact potential in (\ref{MasterEquation}). The plots use the same numerical values of $\gamma$ as in Figure~\ref{Fig:CompEpsilonDirect}.}
\label{Fig:QNMPotentialModels}
\end{figure}

We start with the Klein-Gordon equation for a massless scalar field $\Phi$ in the background of the black hole metric (\ref{FormMetric})
\begin{align}
&\nabla_\mu\nabla^\mu \Phi=\frac{1}{\sqrt{-g}}\,\partial_\mu\left(\sqrt{-g}\,g^{\mu\nu}\,\partial_\nu\Phi\right)=0\,,&&\text{with} &&\Phi(t,z,\theta,\varphi)=\sum_{\ell,m}\frac{\scal_{\ell m}(t,r)}{Mr}\,Y_{\ell m}(\theta,\varphi)\,,\label{KGequation}
\end{align}
where $(t,z,\theta,\varphi)$ are the same coordinates we have used to define the metric $g_{\mu\nu}$ in (\ref{FormMetric}), while $r$ is defined as in (\ref{DimensionlessQuant}). Upon further using the Fourier transform
\begin{align}
\scal_{\ell m}(t,r)=\frac{1}{\sqrt{2\pi}}\int_{\mathbb{R}}d\omega\,\widehat{\scal}_{\ell m}(\omega,r)\,e^{-i\omega t/M}\,,
\end{align} 
we can reduce (\ref{KGequation}) to the radial equation \cite{Nollert:1999ji}
\begin{align}
&\frac{d^2\widehat{\scal}_{\ell m}}{dr_*^2}+\left(\omega^2-V_{\text{scal}}(r)\right)\,\widehat{\scal}_{\ell m}=0\,,&&\text{with} &&V_{\text{scal}}(r)=f\,\left(\frac{\ell(\ell+1)}{r^2}+\frac{1}{r}\,\frac{df}{dr}\right)\,.\label{MasterEquation}
\end{align}
Here we have defined the tortoise coordinate $r_*$ through $\frac{dr_*}{d\tco}=\frac{\sqrt{1-f(\tco)}}{f(\tco)}$ (or equivalently $\frac{dr_*}{dr}=\frac{1}{f(r)}$), which is approximated by\footnote{While it is possible to also expand $r_*$ itself within the linear response formalism, it is not necessary for the subsequent discussion.}
\begin{align}
&\frac{dr_*}{d\tco}\sim\frac{2\sh^{1/3}}{\sh^{2/3}-4}+\coff{\tfrac{dr_*}{d\tco}}{\con}(\tco)\,\con+\sum_{k=1}^2\coff{\tfrac{dr_*}{d\tco}}{k}(\tco)\,\mathcal{I}_{k}[\eps](\tco)\,,&&\text{with} &&\begin{array}{l}\coff{\tfrac{dr_*}{d\tco}}{\con}(\tco)=\frac{2\tco(\sh^{4/3}-16)}{\sh^{2/3}(\sh^{2/3}-4)^3}\,, \\[4pt] 
\coff{\tfrac{dr_*}{d\tco}}{1}(\tco)=\frac{4\sh^{1/3}+\sh}{2(\sh^{2/3}-4)^2}\,,\\[4pt]
\coff{\tfrac{dr_*}{d\tco}}{2}(\tco)=-\frac{4\sh^{1/3}+\sh}{2(\sh^{2/3}-4)^2}\,.
\end{array}
\end{align}
As before (see (\ref{ShorthandQnotation})), In order to keep the notation compact, we have used $\sh=8+6\tco$. Furthermore, we can rewrite the potential $V_{\text{scal}}$ in (\ref{KGequation}) as a function of $\tco$, for which we find 
\begin{align}
V_{\text{scal}}(\tco)\sim\frac{4(\sh^{2/3}-4)(4+\ell(\ell+1)\sh^{2/3})}{\sh^{8/3}}+\coff{V_{\text{scal}}}{\con}(\tco)\,\con+\sum_{k=0}^2\coff{V_{\text{scal}}}{k}(\tco)\,\mathcal{I}_{k}[\eps](\tco)\,,\label{VscalLin}
\end{align}
with
\begin{align}
&\coff{V_{\text{scal}}}{\con}(\tco)=\frac{2^{11}}{3\sh^{11/3}}-\frac{4}{3\sh^3}\,\left(16(6+\sh^{1/3})+\ell(\ell+1)(\sh^{2/3}(16+\sh)-96)\right)\,,\nonumber\\
&\coff{V_{\text{scal}}}{0}(\tco)=\frac{4}{\sh^{2/3}}-1\,,\hspace{2cm}\coff{V_{\text{scal}}}{1}(\tco)=-\frac{8}{\sh^{8/3}}\left(8+(\ell(\ell+1)-1)\sh^{2/3}\right)\,,\nonumber\\%-\frac{4}{\sh^{8/3}}\left(12+\left(2\ell(\ell+1)-1\right)\sh^{2/3}+4\sh-\sh^{5/3}\right)\,,\nonumber\\
&\coff{V_{\text{scal}}}{2}(\tco)=\frac{2^7}{\sh^{8/3}}-\frac{4}{\sh^2}\left(6+\ell(\ell+1)(\sh^{2/3}-6)\right)\,.
\end{align}
In order to illustrate (\ref{VscalLin}), we have plotted $V_\scal$ in Figure~\ref{Fig:QNMPotentialModels} for the three examples introduced in Section~\ref{Sect:ExamplesMetrics}: the plots show the linear response (solid lines) in comparison to the exact (numerical) potentials for a fixed (small) value of $\gamma$. As is evident from these plots, the linear approach provides potentials that are very close to the exact potentials in the deformed space-times.

%%%%%%%%%%%%%%%%%%%%%%%

\subsection{Scalar Quasi-normal Modes}\label{Sect:ScalarQNM}
\begin{figure}[h]
\begin{center}
\includegraphics[width=7.5cm]{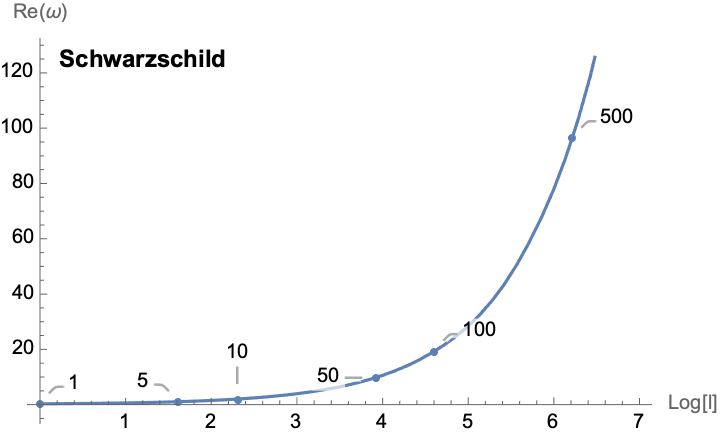}\hspace{1cm}\includegraphics[width=7.5cm]{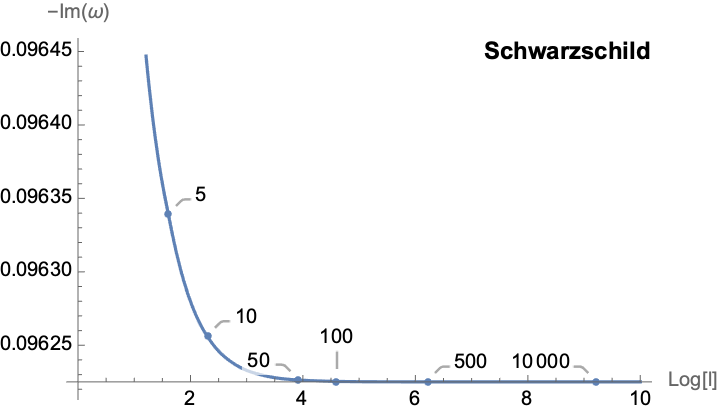}
\end{center}
\caption{Real (left panel) and imaginary (right panel) part of the scalar QNM for the (undeformed) Schwarzschild black hole with $n=0$, as a function of $\ell$. The labeled points indicate specific values of $\ell$, which have been obtained using a third order WKB approach in (\ref{QNMWKB}). The solid line represents an (interpolating) function for generic $\ell\in\mathbb{R}_+$.}
\label{Fig:WKB_direct_QNM_Schwarz}
\end{figure}

In order to solve (\ref{MasterEquation}), even for the linear approximation of the potential (\ref{VscalLin}), still requires to specify suitable boundary conditions. In this Subsection, we shall discuss \emph{quasi-normal modes} (QNM)~\cite{Kokkotas:1999bd,Nollert:1999ji,Berti:2009kk,Konoplya:2011qq}, \emph{i.e.} solutions with purely outgoing boundary conditions 
\begin{align}
&\widehat{\phi}_{\ell m}\sim_{r_*\to\infty}e^{i\omega r_*}\,,&&\text{and} &&\widehat{\phi}_{\ell m}\sim_{r_*\to-\infty}e^{-i\omega r_*}\,,
\end{align}
In the context of black holes, this leads to solution representing characteristic eigenfrequencies \cite{Regge:1957td} (see also \cite{Zerilli:1970wzz,Zerilli:1970se}) that can be used to approximate the spectra of gravitational waves observed in black hole merger events \cite{Nollert:1999ji}. Due to their physical relevance, over the years various different (numerical) techniques have been developed to compute the spectrum of QNM for a given geometry (\emph{e.g.} \cite{Leaver:1985ax,Schutz:1985km,Iyer:1986np,Iyer:1986nq,Cho:2009cj,Cho:2011sf,Chen:2017kqa,Cardoso:2019mqo,,McManus:2019ulj,Livine:2024bvo}. Since different methods introduce their own (numerical) errors in this calculated spectrum, we shall not contrast different such methods here, but only consider one in particular.\footnote{A more systematic discussion of the error arising due to the linear response approach, relative to the numerical error of using different methods to compute the QNM spectrum, shall be presented elsewhere.} Indeed, here we shall compute the spectrum using a third-order WKB approach, as outlined in \cite{Iyer:1986np,Iyer:1986nq} (generalising the second order result of \cite{Schutz:1985km}).\footnote{The WKB approach was further generalised to higher orders \cite{Konoplya:2003ii,Matyjasek:2017psv,Konoplya:2019hlu,Miyachi:2025ptm,Matyjasek:2026yiu}. While our approach can be directly generalised, here, for simplicity, we limit ourselves to third order.} This WKB approximation generally works well in the \emph{eikonal} regime $\ell\gg 1$ and for small overtone index $n\ll\ell$, in which case we have
\begin{align}
\omega^2=\left(\scalv+(-2\scalv^{(2)})^{1/2}\Lambda(n)\right)-i\left(n+\tfrac{1}{2}\right)(-2\scalv^{(2)})^{1/2}\,(1+\Omega(n))\,,\label{QNMWKB}
\end{align}
with
{\allowdisplaybreaks
\begin{align}
\Lambda(n)&=\frac{1}{(-2\scalv)^{1/2}}\left[\frac{1}{8}\left(\frac{\scalv^{(4)}}{\scalv^{(2)}}\right)\left(\frac{1}{4}+(n+\tfrac{1}{2})^2\right)-\frac{1}{288}\left(\frac{\scalv^{(3)}}{\scalv^{(2)}}\right)^2(7+60(n+\tfrac{1}{2})^2)\right]\,,\nonumber\\
\Omega(n)&=\frac{1}{(-2\scalv^{(2)})}\bigg[\frac{5}{6912}\left(\frac{\scalv^{(3)}}{\scalv^{(2)}}\right)^4(77+188(n+\tfrac{1}{2})^2)-\frac{1}{384}\frac{(\scalv^{(3)})^2\scalv^{(4)}}{(\scalv^{(2)})^3}(51+100(n+\tfrac{1}{2})^2)\nonumber\\
&\hspace{1cm}+\frac{1}{2304}\left(\frac{\scalv^{(4)}}{\scalv^{(2)}}\right)^2(67+68(n+\tfrac{1}{2})^2)+\frac{1}{288}\frac{\scalv^{(3)}\scalv^{(5)}}{(\scalv^{(2)})^2}(19+28(n+\tfrac{1}{2})^2)\nonumber\\
&\hspace{1cm}-\frac{1}{288}\frac{\scalv^{(6)}}{\scalv^{(2)}}(5+4(n+\tfrac{1}{2})^2)\bigg]\,.
\end{align}}
Here $\scalv^{(k)}=\frac{d^{k} V_{\text{scal}}}{dr_*^{k}}(\tco_{\text{max}})$ denotes the $k$-th derivative of the potential $V_{\text{scal}}$ in (\ref{MasterEquation}) with respect to the tortoise-coordinate $r_*$, evaluated at the maximum $\tco_{\text{max}}$ of $V_{\text{scal}}$ \cite{Iyer:1986vv,Iyer:1986nq}. For $\con=0=\eps$, this maximum of the potential is at
\begin{align}
\tco_{\text{max},0}=&\frac{1}{3}\bigg(-4+\frac{1}{\ell^{3/2}(\ell+1)^{3/2}}\bigg(9(3\ell^6+9\ell^5+8\ell^4+\ell^3+\ell-3)\nonumber\\
&\hspace{1cm}+\sqrt{(9+\ell(\ell+1)(3\ell-2)(3\ell+5))^2(9+\ell(\ell+1)(14+9\ell(\ell+1)))}\bigg)^{1/2}\bigg)\,,\label{FunctionTmax}
\end{align}
and the real- and imaginary part of the frequencies (\ref{QNMWKB}) are shown in Figure~\ref{Fig:WKB_direct_QNM_Schwarz}.
Notice in particular the asymptotic behaviour for $\ell\to\infty$
\begin{align}
&\text{Re}(\omega)\big|_{\con=0,\eps=0}\sim_{\ell\to\infty} \frac{\ell}{3\sqrt{3}}\,,&&\text{Im}(\omega)\big|_{\con=0,\eps=0}\sim_{\ell\to\infty}-\frac{\left(n+\frac{1}{2}\right)}{3\sqrt{3}}\,.
\end{align}
\begin{figure}[h]
\begin{center}
\includegraphics[width=7.5cm]{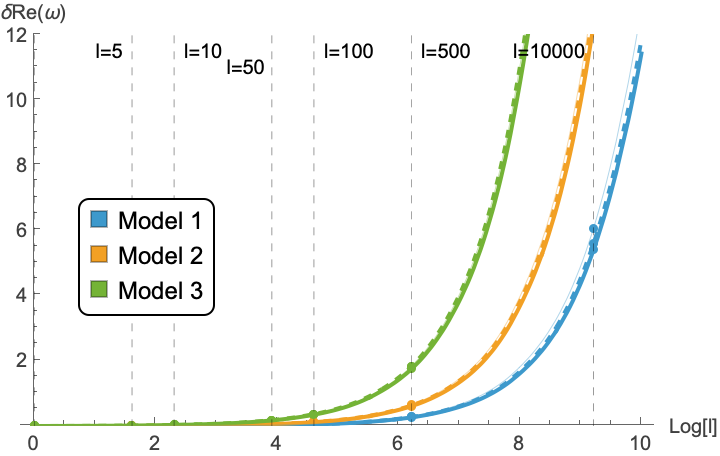}\hspace{1cm}\includegraphics[width=7.5cm]{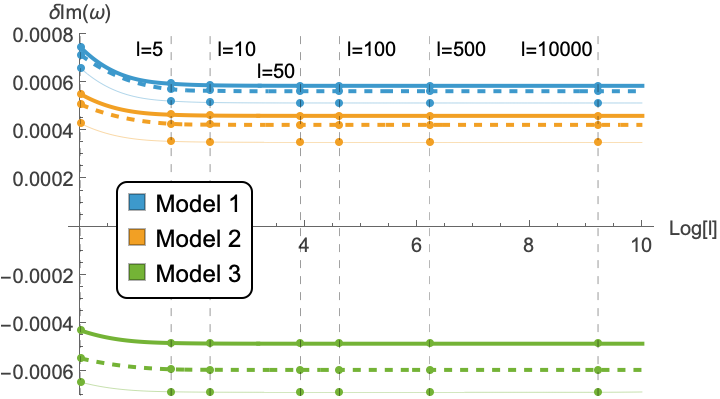}
\end{center}
\caption{Real (left panel) and imaginary part (right panel) of the linear correction $\delta\omega(\ell,n):=\omega(\ell,n)-\omega(\ell,n)\big|_{\con=0,\eps=0}$ to the scalar quasi normal modes for $n=0$, as a function of $\ell$ in the case of the three examples introduced in Section~\ref{Sect:ExamplesMetrics}. The frequencies have been calculated using the third order WKB approach: in both cases the dashed line represents the exact numerical results (within the framework of the WKB approach), the solid heavy line shows the linear response (\ref{LinearResponseQNM}) with the exact momenta $\mathcal{I}_n[\eps]$ calculated numerically, while the thin solid line represents the linear response (\ref{LinearResponseQNM}) with the momenta $\mathcal{I}_n[\eps]$ as in Section~\ref{App:MomentaExamples}. The values of the deformation parameter $\gamma$ for the three examples in the right panel, is the same as in Figure~\ref{Fig:CompEpsilonDirect}.
}
\label{Fig:WKB_Linear_QNM}
\end{figure}
Corrections to the undeformed case shown in Figure~\ref{Fig:WKB_Linear_QNM} can be calculated using the linear response approach
\begin{align}
&\omega(\ell,n)\sim \omega(\ell,n)\big|_{\con=0,\eps=0}+\coff{\omega}{\con}(\ell,n)\,\con+\sum_{k=-6}^2\coff{\omega}{k}(\ell,n)\,\mathcal{I}_k[\eps](\tco_{\text{max},0})\,,\label{LinearResponseQNM}
\end{align}
where the coefficients $\coff{\omega}{\con}$ and $\coff{\omega}{k}$ can be computed explicitly: in Figure~\ref{Fig:CompEpsilonDirect} we have plotted the corrections (\emph{i.e.} $\delta\omega(\ell,n):=\omega(\ell,n)-\omega(\ell,n)\big|_{\con=0,\eps=0}$) for $n=0$ and different values of $\ell$ for three models of black holes.

\begin{figure}[h]
\begin{center}
\includegraphics[width=10cm]{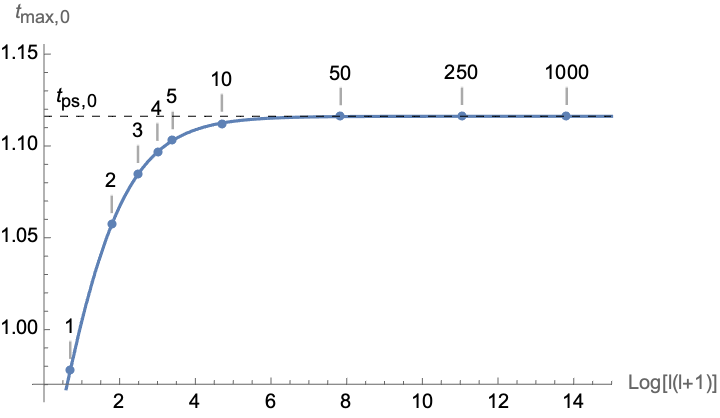}
\end{center}
\caption{$\tco_{\text{max},0}$ as a function of $\ell$ (for convenience, we are using a logarithmic scale). The points indicate quantised values of $\ell$ (with the value of $\ell$ indicated), while the solid blue line shows (\ref{FunctionTmax}) for continuous $\ell$. The dashed black line indicates the asymptotic value $\tpo=\sqrt{6}-\tfrac{4}{3}$, which is the classical position of the photon sphere. Notice that for $\ell\gg 1$, the points become dense, thus justifying to consider (\ref{FunctionTmax}) as a continuous function.}
\label{Fig:PosMax}
\end{figure}

Graphically, the relation between $\tco_{\text{max},0}$ and $\ell$ is shown in Figure~\ref{Fig:PosMax}
and asymptotically, we find
\begin{align}
\lim_{\ell\to \infty}\tco_{\text{max},0}=\tco_{\text{ps}}\big|_{\con=0,\eps=0}=\tco_{\text{ps},0}=\frac{3\sqrt{6}-4}{3}\,,
\end{align}
which is the positon of the photon sphere (see eq.~(\ref{PositionPhotonSphere})) in the undeformed case. In other words, for $\ell\in\mathbb{N}^*$
\begin{align}
&\tco_{\text{max},0}(\ell)\in\left[\frac{1}{3}\left(-4+\frac{1}{2\sqrt{2}}\,\sqrt{171+25\sqrt{73}}\right),\frac{3\sqrt{6}-4}{3}\right]\,,&&\text{with} &&\tco_{\text{max},0}(\ell)<\tco_{\text{max},0}(\ell+1)\,,
\end{align}
is a sequence of points which converges to $\tco_{\text{ps},0}=\frac{3\sqrt{6}-4}{3}$. In fact, for large enough $\ell$, \emph{i.e.} in the eikonal regime, the $\tco_{\text{max},0}(\ell)$ form a quasi-continuous parametrisation of the region close to (but smaller than) the (undeformed) photon sphere. Indeed, (\ref{FunctionTmax}) can be inverted
\begin{align}
\ell(\ell+1)=\frac{96\cdot 2^{1/2}-4\cdot 2^{2/3}(4+3\tco)^{1/3}-3(9\cdot 2^{1/3}\tco^2+2(4+3\tco)^{2/3}+2^{1/3}\tco(24+(8+6\tco)^{1/3}))}{(4+3\tco)^{2/3}(3\tco(8+3\tco)-38)}
\end{align}
which holds for $\tco=\tco_{\text{max},0}$. In the following, we shall extend $\tco$ in this relation to a continuous variable $\tco\leq \frac{3\sqrt{6}-4}{3}=\tco_{\text{ps},0}$ that parametrises space-time close to the photon sphere. This corresponds to a continuous $\ell\gg 1$ parametrising the eikonal regime. 

In this way, we can the re-write the QNM frequencies in the linear response approach as functions of a continuous variable
\begin{align}
&\omega(\tco)\sim \omega(\tco)\big|_{\con=0,\eps=0}+\coff{\omega}{\con}(\tco)\,\con+\sum_{k=-6}^2\coff{\omega}{k}(\tco)\,\mathcal{I}_k[\eps](\tco)\,,&&\text{for} &&\tco\leq \frac{3\sqrt{6}-4}{3}=\tco_{\text{ps},0}\,,\label{LinearResponseQNMF}
\end{align}
where $\omega(\tco)\big|_{\con=0,\eps=0}$ is given by the spectrum of QNMs of the Schwarzschild black hole (see Figure~\ref{Fig:WKB_direct_QNM_Schwarz}). For better readability, we have suppressed the argument $n$ in the coefficients $\coff{\omega}{\con}(\tco)$ and $\coff{\omega}{k}(\tco)$. We can provide explicit expressions for the leading expansion of thes coefficients around $\tpo=\sqrt{6}-\tfrac{4}{3}$, \emph{i.e.} the (classical) localisation of the photon sphere
{\allowdisplaybreaks
\begin{align}
&\text{Re}(\coff{\omega}{\con}(\tco))=\text{Re}(\coff{\omega}{1}(\tco))=-\frac{(\tpo-\tco)^{-1/2}}{6\cdot 54^{1/4}}+\frac{5(6n(n+1)+13)}{648\cdot 24^{1/4}}\,(\tpo-\tco)^{1/2}+\mathfrak{o}((\tpo-\tco)^{1/2})\,,\nonumber\\
&\text{Re}(\coff{\omega}{0}(\tco))=\frac{30\,n(n+1)-97}{36\cdot 24^{1/4}}\,(\tpo-\tco)^{1/2}+\mathfrak{o}((\tpo-\tco)^{1/2})\,,\nonumber\\
&\text{Re}(\coff{\omega}{-1}(\tco))=\frac{2310\,n(n+1)+499}{256\cdot 24^{1/4}}\,(\tpo-\tco)^{1/2}+\mathfrak{o}((\tpo-\tco)^{1/2})\,,\nonumber\\
&\text{Re}(\coff{\omega}{-2}(\tco))=\frac{3^{3/4}(510\,n(n+1)+223)}{128\cdot 2^{3/4}}\,(\tpo-\tco)^{1/2}+\mathfrak{o}((\tpo-\tco)^{1/2})\,,\nonumber\\
&\text{Re}(\coff{\omega}{-3}(\tco))=\frac{81\cdot 3^{1/4}(2\,n(n+1)+1)}{128\cdot 2^{1/4}}\,(\tpo-\tco)^{1/2}+\mathfrak{o}((\tpo-\tco)^{1/2})\,,\nonumber\\
&\text{Re}(\coff{\omega}{2,-4,-5,-6}(\tco))=\mathfrak{o}((\tpo-\tco)^{1/2})\label{ReQNM}
\end{align}}
and 
{\allowdisplaybreaks
\begin{align}
&\text{Im}(\coff{\omega}{\con}(\tco))=\text{Im}(\coff{\omega}{1}(\tco)=\frac{n+\tfrac{1}{2}}{6\sqrt{3}}+\frac{(n+\tfrac{1}{2})\left(235n(n+1)+137\right)}{11664\sqrt{2}}\,(\tpo-\tco)+\mathfrak{o}(\tpo-\tco)\,,\nonumber\\
&\text{Im}(\coff{\omega}{0}(\tco))=\frac{5(n+\tfrac{1}{2})\left(47n(n+1)-122\right)}{324\sqrt{2}}\,(\tpo-\tco)+\mathfrak{o}(\tpo-\tco)\,,\nonumber\\
&\text{Im}(\coff{\omega}{-1}(\tco))=-\frac{9(n+\tfrac{1}{2})}{4\sqrt{2}}+\frac{(n+\tfrac{1}{2})\left(90170n(n+1)-58629\right)}{6144\sqrt{2}}\,(\tpo-\tco)+\mathfrak{o}(\tpo-\tco)\,,\nonumber\\
&\text{Im}(\coff{\omega}{-2}(\tco))=\frac{(n+\tfrac{1}{2})\left(109030n(n+1)+43109\right)}{3072\sqrt{2}}\,(\tpo-\tco)+\mathfrak{o}(\tpo-\tco)\,,\nonumber\\
&\text{Im}(\coff{\omega}{-3}(\tco))=\frac{3\sqrt{3}(n+\tfrac{1}{2})\left(4098n(n+1)+4031\right)}{1024}\,(\tpo-\tco)+\mathfrak{o}(\tpo-\tco)\,,\nonumber\\
&\text{Im}(\coff{\omega}{-4}(\tco))=\frac{27(n+\tfrac{1}{2})\left(82n(n+1)+107\right)}{256\sqrt{2}}\,(\tpo-\tco)+\mathfrak{o}(\tpo-\tco)\,,\nonumber\\
&\text{Im}(\coff{\omega}{-5}(\tco))=\frac{81\sqrt{3}(n+\tfrac{1}{2})\left(2n(n+1)+3\right)}{512}\,(\tpo-\tco)+\mathfrak{o}(\tpo-\tco)\,,\nonumber\\
&\text{Im}(\coff{\omega}{2,-6}(\tco))=\mathfrak{o}(\tpo-\tco)\label{ImQNM}
\end{align}}
where $\mathfrak{o}$ denotes subleading orders, which can be neglected in the eikonal regime. We remark that in the eikonal limit $\ell\to\infty$ (\emph{i.e.} $\tco\to \tco_{\text{ps},0}$) we find 
\begin{align}
&\text{Re}(\coff{\omega}{\con}(\tco))=\text{Re}(\coff{\omega}{1}(\tco))\sim_{\ell\to\infty}-\frac{\ell}{6\sqrt{3}}\,,\nonumber\\
&\text{Im}(\coff{\omega}{\con}(\tco))=\text{Im}(\coff{\omega}{1}(\tco)\sim_{\tco\to\tco_{\text{ps},0}}\frac{n+\tfrac{1}{2}}{6\sqrt{3}}\,,&&\text{Im}(\coff{\omega}{-1}(\tco))\sim_{\tco\to\tco_{\text{ps},0}}-\frac{9(n+\tfrac{1}{2})}{4\sqrt{2}}\,,
\end{align}
while
\begin{align}
&\text{Re}(\coff{\omega}{k}(\tco))\sim_{\ell\to\infty}0\hspace{0.4cm}\forall k\neq1\,,&&\text{and}&&\text{Im}(\coff{\omega}{k}(\tco))\sim_{\tco\to\tco_{\text{ps},0}}0\hspace{0.4cm}\forall k\neq\pm1\,,
\end{align}
Together with (\ref{LinearResponseQNM}) as well as (\ref{Mogc}) and (\ref{Lyapunov}) we therefore find
\begin{align}
\omega\sim_{\ell\to \infty}M\,\ell\,\omega_c-\left(n+\tfrac{1}{2}\right)\,|\lambda|\,,\label{EikonalLimit}
\end{align}
which was already observed in \cite{Cardoso:2008bp}. The linear response relation (\ref{LinearResponseQNMF}) therefore generalises (\ref{EikonalLimit}) beyond the \emph{eikonal limit} to a region for large (but not infinite) $\ell$. Alternatively, when interpreting the (eikonal) spectrum of QNMs as physical input, (\ref{LinearResponseQNMF}) can be understood as an integro-differential equation for $\eps$.

%%%%%%%%%%%%%%%%%%%%%%%%%%%%%
%%%%%%%%%%%%%%%%%%%%%%%%%%%%%
%%%%%%%%%%%%%%%%%%%%%%%%%%%%%
\section{Conclusions}\label{Sect:Conclusions}
In this work, we have proposed a novel effective description of a class of spherically symmetric and static black holes that allows to extract universal features of physics beyond (vacuum) GR from observables of the geometry.  We describe deformations of the black hole space-time beyond the Schwarzschild geometry by introducing a function $\eps$ at every point of space-time outside of the horizon, which we interpret as the single eigenvalue of the Einstein tensor. Since the latter is identically zero for the Schwarzschild black hole, $\eps$ encodes deformations of the latter. In fact, inspired by the linear response approach to perturbative problems in physics, interpreting $\eps$ as an \emph{external} perturbation of the classical geometry, we have developed a framework to write the leading corrections of general observables computed from this geometry. Concretely, a generic observable $\mathbb{O}$  can be approximated in the form of the expansion (\ref{LinearResponse}), where $\mathcal{I}_n[\eps](\tco)$ are linear functionals of $\eps(\tco)$ (\emph{i.e.} derivatives or (nested) integrals, as specified in (\ref{Momenta})) and $\con$ is a (small) constant parameter (chosen here to be the normalised deviation of the position of the event horizon), which is required as further input for certain observables. $\eps$ and $\mathcal{I}_n[\eps]$ are defined at every point in space-time and as a concrete parametrisation, we have chosen in this work $\tco\in[0,\infty)$, the proper time of a radially in-falling massive observer to reach the horizon (normalised by the mass of the black hole), as defined in (\ref{DefTau}). The choice of this parameter does not change the structure of (\ref{LinearResponse}), but only changes the coefficients $\{\coff{\mathbb{O}}{\con}\,,\coff{\mathbb{O}}{n}(\tco)\}$ in a way that only depends on the undeformed geometry, as explained in Section~\ref{Sect:VariableChange}.

The momenta $\mathcal{I}_n[\eps](\tco)$ and $\con$ in (\ref{LinearResponse}) are model dependent, \emph{i.e.} they are implicitly specified by a theory of gravity beyond the GR benchmark or determined in an ad hoc way through other theoretical modelling. A number of examples of $\eps$ are plotted schematically in Figure~\ref{Fig:CompEpsilonDirect} and some of its momenta in Figure~\ref{Fig:MomentaModels}. The coefficients $\{\coff{\mathbb{O}}{\con}\,,\coff{\mathbb{O}}{n}(\tco)\}$ are model independent (\emph{i.e.} independent of $\eps$ and $\con$) and are specific to the observable $\mathbb{O}$. They therefore encode universal features of physics beyond GR that is encoded in the latter. We have concretely worked out these coefficients for a number of observables, notably
\begin{itemize}
\item[\emph{(i)}] the surface gravity $\kappa$ (see (\ref{Fig:Temp_Examples})), which depends on momenta $\mathcal{I}_n[\eps]$ evaluated at the event horizon (\emph{i.e.} at $\tco=0$)
\item[\emph{(ii)}] quantities related to the geodesic motion of massless test particles, namely the position of the photon sphere $\tco_{\text{ps}}$ (see (\ref{PositionPhotonSphere})), the black hole shadow (see (\ref{BHshadow}) and also \cite{DelPiano:2024nrl}), the frequency of photons circling the black hole at the photon sphere (see (\ref{Mogc})) and the Lyapunov exponent (see (\ref{Lyapunov})) which is related to the stability of the photon sphere. For all these observables, the momenta $\mathcal{I}_n[\eps]$ are evaluated at the position of the photon sphere $\tco_{\text{ps},0}$ (of the undeformed geometry). 
\item[\emph{(iii)}] quantities related to the geodesic motion of massive test particles, namely the position of the ISCO (innermost stable circular orbit, see (\ref{Tisco})) and the critical angular momentum (see (\ref{CriticalAngular})). Both observables depend on momenta $\mathcal{I}_n[\eps]$ evaluated at the (undeformed) position of the ISCO ($\tco=\frac{4(3\sqrt{3}-1)}{3}$)
\item[\emph{(iv)}] the frequency-spectrum of quasi-normal modes (QNMs) of massless scalar fields in the background of the black hole geometry (see (\ref{ReQNM}) for the real part and (\ref{ImQNM}) for the imaginary part), approximated through a third order WKB method. These frequencies depend on momenta $\mathcal{I}_n[\eps]$ evaluated at the maximum of the potential (\ref{VscalLin}), which depends on the mode number $\ell$. In the eikonal region $\ell\gg 1$, the latter becomes quasi-continues, such that the QNM frequencies depend in momenta evaluated in a region close to (but inferior to) the (undeformed) position of the photon sphere $\tco_{\text{ps},0}$
\end{itemize}
We have systematically illustrated all observables at the hand of three different examples (\emph{i.e.} three different choices of functions for $\eps$ and constants $\con$), as explained in detail in Section~\ref{Sect:ExamplesMetrics}. Moreover, taking the values of these observables as input (\emph{e.g.} as the result of some measurement), the universality of the coefficients $\{\coff{\mathbb{O}}{\con}\,,\coff{\mathbb{O}}{n}\}$ allows to directly establish non-trivial relations for the functional dependence of $\eps$ on the space-time position. Interpreting the latter as the energy density that enters as (external) source into the Einstein equation (as field equation for the black hole geometry), we therefore obtain direct input for an effective description of the black hole. As schematically shown in Figure~\ref{Fig:LampPost}, this furnishes a description of effective physics (concretely an external source) beyond the GR benchmark that gives rise locally to the black hole space-time in a region outside of the horizon.

The observables discussed in this paper are only examples, which were selected to demonstrate the viability (and versatility) of our approach. In the future we plan to further extend the above list, for example by considering QNMs of other fields than massless scalars. This will furnish further integro-differential equations for $\eps$ in certain regions of space-time.

A simplification of our approach is the limitation to spherically symmetric and static black holes, whose metric is of the form (\ref{FormMetric}). For this class, the Einstein tensor has only one functionally independent eigenvalue, thus simplifying our discussion. For the future, we plan to generalise the discussion by allowing for a second eigenvalue (which is also assumed to be small). It would be interesting to extend the approach further to also include black hole geometries with angular momentum.

Finally, while our approach allows to directly relate observables to effective descriptions of black holes, it is per se not a fundamental approach to gravity beyond GR. Indeed, while a (quantum) theory of gravity would predict the functional $\mathbb{O}[\con,\eps]$ in (\ref{LinearResponse}), such a theory would be required to do more: indeed in a full theory beyond vacuum GR, both $\con$ and $\eps$ should not act as external sources, but rather should be determined through solving some intricate set of field equations for the metric and potential matter. Nevertheless, our approach furnishes an important intermediate step towards such a theory, as it allows to relate information from (measurable) observables into information about partial solutions to this system of equations: concretely, for any observable, assuming the deviation from GR to be small, it allows to separate aspects that are independent of $\eps$ and $\con$ (which we termed model-independent) and provides a functional form in which effects beyond the GR benchmark enter.
%%%%%%%%%%%%%%%%%%%%%%%%%%%%%
%%%%%%%%%%%%%%%%%%%%%%%%%%%%%
%%%%%%%%%%%%%%%%%%%%%%%%%%%%%
\section*{Acknowledgements}
I am very grateful to Zakaria Belkhadria, Anna Chrysostomou, Aldo Deandrea, Manuel Del Piano, Luciano Rezzolla, Hugo Roussille and Francesco Sannino for numerous discussions and exchanges on related topics. I am also thankful to the Quantum Theory Center of the University of Southern Denmark for kind hospitality, while part of this work was being completed.

\appendix
%%%%%%%%%%%%%%%%%%%%%%%%%%%%%%%%%%%%%%%%%%%%
\section{Geometry and Linear Response}

\subsection{Review of Series Expansion}\label{App:SeriesExpansion}
In order to complement the discussion of Section~\ref{Sect:LinRespsone}, we briefly review the results of \cite{SH}. In the current notation, \cite{SH} considered a series expansion of the function $\Phi$ in the geometry (\ref{FormMetric})
\begin{align}
\Phi=M(2+\con)+\sum_{n=1}^\infty \frac{v_n}{(2M)^{n-1}}\,\tau^n\,,\label{SeriesExpansion}
\end{align}
with $\tau$ defined in (\ref{DefTau}) and $(v_n)_{n\in\mathbb{N}}$ a set of small expansion coefficients, which parametrise the deviation from the Schwarzschild geometry. In \cite{SH} it was assumed that $\Phi$ is sufficiently regular to allow for an expansion (\ref{SeriesExpansion}), which converges at least in a region close enough to the horizon. The eigenvalue $\epsilon$ can be similarly expanded
\begin{align}
&\epsilon(\tau)=\sum_{m,n=1}^\infty \mathcal{M}_{nm}\,\frac{\tau^n v_m}{(2M)^{n+2}}+\mathcal{O}(v_n^2)\,,&&\text{with} &&\mathcal{M}=\left(\begin{linesmall}{@{}ccccc@{}}&&&\\[-6pt]1 & 0 & 0 & 0 & \cdots \\[3pt]-\tfrac{3}{2} & 2 & 0 & 0 & \cdots \\[3pt] \tfrac{9}{4} & -3 & 3 & 0 & \cdots \\[3pt] -\frac{27}{8} & \tfrac{9}{2} & -\tfrac{9}{2} & 4 & \cdots \\[3pt] \vdots & \vdots & \vdots & \vdots & \ddots\end{linesmall}\right)\,.\label{ExpansionEps}
\end{align}
Here the corrections $\mathcal{O}(v_n^2)$ are assumed to be small and neglected. (\ref{ExpansionEps}) can be resummed
\begin{align}
\epsilon(\tau)&=\sum_{m=1}^\infty\frac{2m\,\tau^{m-1}\,v_m}{(3\tau+4M)(2M)^m}+\mathcal{O}(v^2)%=\frac{1}{(3\tau+4M)\,M}\,\frac{d}{d\tau}\sum_{m=1}^\infty\frac{v_m}{(2M)^{m-1}}\,\tau^m+\mathcal{O}(v^2)\nonumber\\
=\frac{1}{(3\tau+4M)\,M}\,\frac{d\Phi}{d\tau}(\tau)+\mathcal{O}(v^2)\,,
\end{align}
which corresponds to (\ref{RelImportEpsPhi}). Similarly, for the radial coordinate as a function of $\tau$, the following expansion was derived
\begin{align}
&z(\tau)=M(2+\con)+\sum_{n=1}^\infty\frac{\tau^n}{(2M)^{n-1}}\left(\frac{3^n\Gamma(5/3)}{2^n\Gamma(n+1)\Gamma(5/3-n)}+\mathcal{A}_n+\sum_{m=1}^\infty\mathcal{B}_{nm} v^m\right)+\mathcal{O}(v_n^2)\,,\label{Zseries}
\end{align}
with the numerical coefficients
\begin{align}
&\mathcal{A}=\left(\begin{linesmall}{@{}c@{}} \\[-6pt]0 \\[3pt] \tfrac{1}{4} \\[3pt] -\tfrac{1}{3} \\[3pt] \tfrac{7}{16} \\[3pt] \vdots \end{linesmall}\right)\,,&&\text{and} &&\mathcal{B}=\left(\begin{linesmall}{@{}ccccc@{}}&&&\\[-6pt]0 & 0 & 0 & 0 & \cdots \\[3pt]\tfrac{1}{4} & 0 & 0 & 0 & \cdots \\[3pt] -\tfrac{1}{8} & \tfrac{1}{6} & 0 & 0 & \cdots \\[3pt] \tfrac{1}{8} & -\tfrac{1}{12} & \tfrac{1}{8} & 0 & \cdots \\[3pt] \vdots & \vdots & \vdots & \vdots & \ddots\end{linesmall}\right)\,.
\end{align}
(\ref{Zseries}) can be resummed in the following form
\begin{align}
z(\tau)=\frac{1}{\left(8+\tfrac{6\tau}{M}\right)^{1/3}}\left((M(2+\con)+2\tau)+\frac{1}{2M}\int_0^\tau dt\,\Phi(t)\right)+\mathcal{O}(v^2)\,.
\end{align}
which corresponds to (\ref{EffectiveProbePosition}). 

The method employed in \cite{SH} to derive (\ref{RelImportEpsPhi}) and (\ref{EffectiveProbePosition}) implicitly relies on series expansions such as (\ref{SeriesExpansion}), which a priori have a finite radius of convergence. However, as functional relations, (\ref{RelImportEpsPhi}) and (\ref{EffectiveProbePosition}) remain true, provided that $\Phi$ as well as its derivatives and integrals remain small.

%%%%%%%%%%%%%%%%%%%%%%%%%%%%%%%%%%%%%%%%%%%%
%%%%%%%%%%%%%%%%%%%%%%%%%%%%%%%%%%%%%%%%%%%%
\subsection{Geodesic Motion}\label{App:GeodesicMotion}
In this Appendix we recall a few salient features of geodesic motion in a spherically symmetric and static metric of the form (\ref{FormMetric}) (see also \cite{Misner:1973prb,Hawking:1973uf,Carroll:2004st,Rezzolla2016}). We shall use dimensionless parameters as in (\ref{DimensionlessQuant}). Our starting point is the position of a test particle $X^\mu(\lambda)$ (with $\lambda$ a suitable parametrisation), which is governed by the geodesic equations 
\begin{align}
&\frac{du^\mu}{d\lambda}+{\Gamma^\mu}_{\nu\rho}u^\nu\,u^\rho=0\,,&&\text{with}&&u^\mu=\dot{X}^\mu:=\frac{dX^\mu}{d\lambda}\,,&&\text{and} &&g_{\mu\nu}u^\mu u^\nu=\zeta\,,\label{GeodesicEq}
\end{align}
Here ${\Gamma^\mu}_{\nu\rho}$ is the connection associated with the metric (\ref{FormMetric}) and $\zeta$ determines the nature of the test particle, \emph{i.e.} in particular $\zeta=0$ for a light-like particle, while $\zeta=-1$ for a space-like particle. The system (\ref{GeodesicEq}) can be re-cast in the following form (see \emph{e.g.} \cite{Carroll:2004st})
\begin{align}
&\frac{d}{d\lambda}\log\left(\dot{t} f\right)=0\,,&&\frac{d}{d\lambda}\log\left(r^2\dot{\varphi}\right)\,,&&-f\,\dot{t}^2+\frac{\dot{r}^2}{f}+r^2\,\dot{\varphi}^2=\zeta\,.\label{RedGeod}
\end{align}
where dots represent derivatives with respect to $\lambda$. The first two equations reveal conserved quantities and can be solved by introducing
\begin{align}
&E=\dot{t}\,f\,,&&\text{and} &&L=r^2\,\dot{\varphi}\,,
\end{align}
where physically $E$ is the energy $E$ and $L$ the angular momentum of the rest particle. Substituting into the remaining equation in (\ref{RedGeod}), we obtain
\begin{align}
&\parbox{4.4cm}{\begin{align}\dot{r}^2&=E^2-f\left(-\zeta+\frac{L^2}{r^2}\right)\nonumber\\
&=E^2-V(r,\zeta)\,,\nonumber
\end{align}}\hspace{0.3cm}\text{with}\hspace{0.2cm}V(r,\zeta)=\left\{\begin{array}{lcl}V_{\text{mass}}(r)=f\,\left(1+\frac{L^2}{r^2}\right) & \text{for} & \zeta=-1\,, \\[4pt] L^2\,V_{\text{massless}}(r)=L^2\,\frac{f}{r^2} & \text{for} & \zeta=0\,.\end{array}\right.\hspace{1.2cm}\label{PotentialRadial}
\end{align}
In this work, we are particularly interested in circular orbits, which are represented by extrema of the potential $V(r,\zeta)$ (see \emph{e.g.} \cite{Rezzolla2016}). Concretely, in the case of the undeformed Schwarzschild black hole ($f(r)=1-\frac{2}{r}$) we have 
\begin{itemize}
\item For a massless particle (\emph{i.e.} $\zeta=0$), $V_{\text{massless}}(r)$ has a single maximum corresponding to the photon sphere at $r_0=3$.
\item For a massive particle (\emph{i.e.} $\zeta=-1$), $V_{\text{mass}}(r)$ generally has a minimum and a maximum, representing stable and unstable circular orbits at $r_0$, depending on the parameter space~\cite{Rezzolla2016}:
\begin{align}
&\text{unstable:}\hspace{0.2cm}\left\{\begin{array}{l}3M\leq r_0\leq 6M\,,\\ 2\sqrt{3}\,M\,\leq L\,,\end{array}\right. &&\text{stable:}\hspace{0.2cm}\left\{\begin{array}{l}6M\leq r_0\,,\\ 2\sqrt{3}\,M\,\leq L\,,\end{array}\right. 
\end{align}
In the extremal case of a single saddle-point of $V_{\text{mass}}$, \emph{i.e.}
\begin{align}
\partial_r V\big|_{r=\ri}=0=\partial^2_r V\big|_{r=\ri}\,.\label{IscoCondition}
\end{align}
 we obtain a single circular orbit, called the innermost stable circular orbit (ISCO) with the parameters $r_{\text{ISCO},0}=6$ and $L=2\sqrt{3}$.
\end{itemize}

%%%%%%%%%%%%%%%%%%%%%%%%%%%%%%%%%%%%%%
%%%%%%%%%%%%%%%%%%%%%%%%%%%%%%%%%%%%%%
%%%%%%%%%%%%%%%%%%%%%%%%%%%%%%%%%%%%%%
\section{Momenta for Examples of Deformed Black Holes}\label{App:MomentaExamples}
\begin{figure}[h]
\begin{center}
\includegraphics[width=5.5cm]{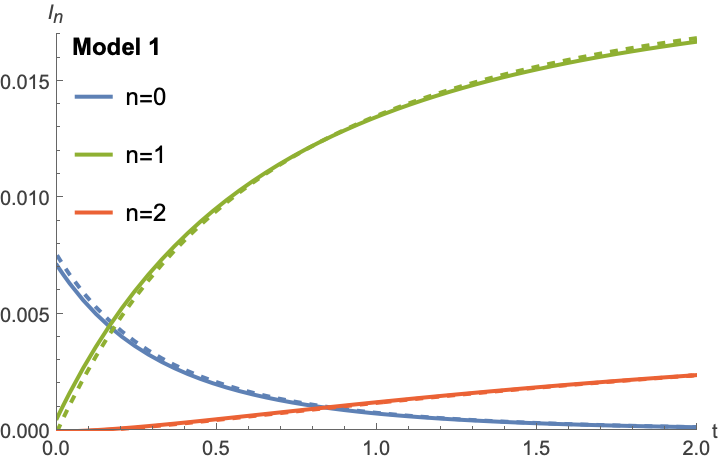}\hspace{0.25cm}\includegraphics[width=5.5cm]{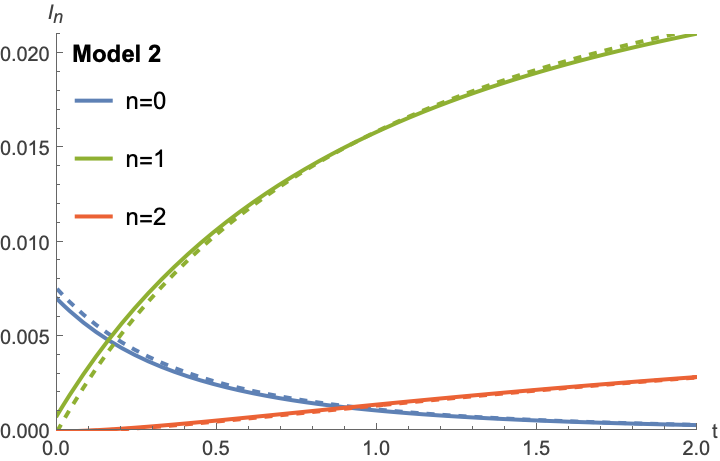}\hspace{0.25cm}\includegraphics[width=5.5cm]{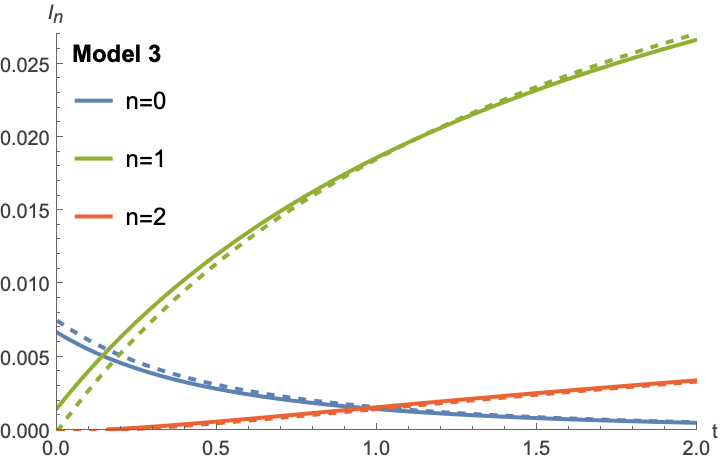}
\end{center}
\caption{The momenta $\mathcal{I}_{n=0,1,2}[\eps](\tco)$ for Model 1 (left panel), Model 2 (middle panel) and Model 3 (right panel) of Table~\ref{Tab:OverviewModels}: the solid lines represent the analytic (linear) approximations (\ref{HaywardMomenta}), (\ref{SV1Momenta}) and (\ref{SV2Momenta}) respectively, while the dashed lines represent numerical computations for the exact momenta. The plots use the same numerical values of $\gamma$ as in Figure~\ref{Fig:CompEpsilonDirect}}
\label{Fig:MomentaModels}
\end{figure}

In this Appendix, we compile (approximate) expressions for the momenta (\ref{Momenta}) for the examples given in Table~\ref{Tab:OverviewModels}. For concreteness, we limit ourselves to the values of $n\in\{0,1,2\}$ that are particularly relevant for the invariants studied in this paper: the $\mathcal{I}_{n>2}[\eps](\tco)$ are not required for the invariants in this work, while the invariants $\mathcal{I}_{n<0}[\eps](\tco)$ can be computed in a straightforward fashion by derivatives. We consider the three examples separately:
\begin{itemize}
\item Model 1: 
The momenta can be approximated in the following form
\begin{align}
&\mathcal{I}_0[\eps](\tco)\sim\frac{192\gamma(1+\gamma)^3}{\left(16+24\tco(1+\gamma)^3+9\tco^2(1+\gamma)^3+16\gamma(4+\gamma(3+\gamma))\right)^2}\,,\nonumber\\
&\mathcal{I}_1[\eps](\tco)\sim \frac{4}{\sqrt{1+\gamma}} \left(\frac{(1+\gamma )^2 (3 \tco+4)}{\sqrt{16 \left(\gamma ^3+3 \gamma ^2+4 \gamma +1\right)+9 (\gamma +1)^3 \tco^2+24 (1+\gamma )^3\tco}}-1\right)\,,\nonumber\\
&\mathcal{I}_2[\eps](\tco)\sim\frac{-4 \sqrt{16 \gamma  (\gamma  (\gamma +3)+4)+9 (\gamma +1)^3 \tco^2+24 (1+\gamma )^3 \tco+16}+12 (\gamma +1)^2 \tco}{3 (\gamma +1)^2 (3 \tco+4)}\nonumber\\
&\hspace{0.5cm}+\frac{16 \left((1+\gamma )^2+\sqrt{\gamma }
   \left(\text{arccoth}\left(\sqrt{\frac{\gamma }{\gamma +\frac{1}{16} (1+\gamma )^3 (3 t+4)^2}}\right)-\text{arctanh}\left(\sqrt{1+\tfrac{1}{\gamma }}\right)\right)\right)}{3 (1+\gamma )^2 (3 \tco+4)}\,.
\label{HaywardMomenta}
\end{align}
which are plotted in the left panel of Figure~\ref{Fig:MomentaModels}. 
%%%%%%%%%%%
\item Model 2: 
{\allowdisplaybreaks
\begin{align}
&\mathcal{I}_0[\eps](\tco)\sim\frac{16\ 2^{2/3} \gamma  (\gamma +1) (\gamma +2)}{(3 \tco+4)^{4/3} \left(4\ 2^{2/3} \gamma 
   (2+\gamma)+(1+\gamma)^2 (3 \tco+4)^{4/3}\right)^{3/2}}\,,\nonumber\\
&\mathcal{I}_1[\eps](\tco)\sim \frac{8\cdot 2^{5/6} \gamma  (2+\gamma ) (3 \tco+4)^{1/3}}{\left(4 (1+\gamma )^2-4\right)
   \left(4-\frac{4}{(\gamma +1)^2}+\frac{(3
   t+4)^{4/3}}{2^{2/3}}\right)^{1/4}}-\frac{4}{\sqrt{\gamma +1}}\nonumber\\
&\mathcal{I}_2[\eps](\tco)\sim \frac{4}{3}-\frac{4}{(1+\gamma )^{3/2}(3\tco+4)}\nonumber\\
&-\frac{2 (\gamma  (2+\gamma))^{1/4} \left((1+\gamma ) (6 \tco+8)^{2/3} \,
   _2F_1\left(\frac{-1}{4},\frac{1}{2};\frac{3}{2};\frac{-(3 \tco+4)^{4/3} (\gamma +1)^2}{4\ 2^{2/3} \gamma  (\gamma +2)}\right)-4 \,
   _2F_1\left(\frac{-1}{4},\frac{1}{2};\frac{3}{2};\frac{-1}{\gamma ^2+2 \gamma }\right)\right)}{(1+\gamma )^2(3 \tco+4)}\,.
\label{SV1Momenta}
\end{align}}
Here $_2F_1(a,b;c;z)$ is a hypergeometric function. The momenta $\mathcal{I}_n[\eps]$ in (\ref{SV1Momenta}) are plotted in the middle panel of Figure~\ref{Fig:MomentaModels}.
%%%%%%%%%%%
\item Model 3: 
{\allowdisplaybreaks
\begin{align}
&\mathcal{I}_0[\eps](\tco)\sim\frac{2^{10/3} (\gamma +1)^{-\frac{2^{4/3}}{(\gamma +1) (3 \tco+4)^{2/3}}-1}\,\log (1+\gamma) }{(3 \tco+4)^{8/3}}\,,\nonumber\\
&\mathcal{I}_1[\eps](\tco)\sim 4 \left((\gamma +1)^{-\frac{2^{1/3}}{(\gamma +1) (3 t+4)^{2/3}}}-\frac{1}{\sqrt{\gamma +1}}\right)\nonumber\\
&\mathcal{I}_2[\eps](\tco)\sim \frac{4 \left(+(1+\gamma)^2 (3 \tco+4) \left(\sqrt{\gamma
   +1}-(1+\gamma)^{\frac{2^{1/3}}{(1+\gamma ) (3 \tco+4)^{2/3}}}\right)\right)}{3 (1+\gamma )^{5/2} (3 \tco+4)}\nonumber\\
&+\frac{4 \left(\log (1+\gamma ) \left(2 \sqrt{1+\gamma} E_{\frac{3}{2}}\left(-\frac{1}{2} \log (1+\gamma)\right)-(1+\gamma ) (6 t+8)^{1/3}
   E_{\frac{3}{2}}\left(-\frac{2^{1/3} \log (1+\gamma)}{(3 \tco+4)^{2/3} (1+\gamma)}\right)\right)\right)}{3 (1+\gamma)^{5/2} (3 t+4)}\
\label{SV2Momenta}
\end{align}}
where $E_n$ is the exponential integral function
\begin{align}
&E_n(x):=\int_1^\infty\frac{e^{-xt}}{t^n}\,dt\,,&&\text{for }n\geq 1\,. 
\end{align}
 The momenta $\mathcal{I}_n[\eps]$ in (\ref{SV2Momenta})  are plotted in the right panel of Figure~\ref{Fig:MomentaModels}.
\end{itemize}
As is indicated in Figure~\ref{Fig:MomentaModels}, in all cases we can also obtain numerical integrals for the momenta, which agree with (the approximations in) (\ref{HaywardMomenta}), (\ref{SV1Momenta}) and (\ref{SV2Momenta}) up to subleading terms.

\noindent
Finally, we remark that in all three examples we find the asymptotic values
\begin{align}
&\lim_{\tco\to\infty} \mathcal{I}_0[\eps](\tco)=0\,,&&\lim_{\tco\to\infty} \mathcal{I}_1[\eps](\tco)\sim4\left(1-\frac{1}{\sqrt{1+\gamma}}\right)\,,&&\lim_{\tco\to\infty} \mathcal{I}_2[\eps](\tco)\sim\frac{4}{3}\left(1-\frac{1}{\sqrt{1+\gamma}}\right)\,,\label{Asymptotic}
\end{align}
which are indeed satisfied by (\ref{HaywardMomenta}), (\ref{SV1Momenta}) and (\ref{SV2Momenta}). The limit of $\mathcal{I}_0[\eps]$ follows from the asymptotic behaviour of the energy density. For $\mathcal{I}_1[\eps]$, we remark that with the approximation (\ref{IntegApprox}), we can write
\begin{align}
\lim_{\tco\to\infty} \mathcal{I}_1[\eps](\tco)\sim \sqrt{2}\int_{\rh}^\infty\frac{\frac{d\fun(\tco(r))}{dr}}{\sqrt{\fun(\tco(r))}}=2\sqrt{2}\,\left[\sqrt{\fun(\tco)}\right]_0^\infty=2\sqrt{2}\left(\sqrt{2}-\sqrt{\frac{2}{1+\gamma}}\right)=4\left(1-\frac{1}{\sqrt{1+\gamma}}\right)\,,\nonumber
\end{align}
where $\fun(\tco)$ is the metric function defined in (\ref{DimensionlessQuant}), with $\fun(0)=\rh=\frac{2}{1+\gamma}$ and $\lim_{\tco\to\infty}\fun(\tco)=2$ for all models. Concerning $\mathcal{I}_2[\eps]$, following the previous argument, we find
\begin{align}
\lim_{\tco\to\infty} \mathcal{I}_2[\eps](\tco)\sim \lim_{\tco\to 0}\frac{2\sqrt{2}}{3\tco+4}\int_0^{\tco}\left(\sqrt{\fun(\tcop)}-\sqrt{\frac{2}{1+\gamma}}\right)d \tcop
\end{align}
From the integral, only the contribution linear in $\tco$ (for $\tco\to\infty$) contributes in the limit, which is independent of the deformation
\begin{align}
\int_0^\tco\left(\sqrt{\fun(\tcop)}-\sqrt{\frac{2}{1+\gamma}}\right)\,d\tcop=\left(\sqrt{2}-\sqrt{\frac{2}{1+\gamma}}\right)\,\tco+\mathfrak{o}(\tco)\,,
\end{align}
which therefore leads to the limit (\ref{Asymptotic}).
%%%%%%%%%%%%%%%%%%%%%%%%%%%%%%%%%%%%%%%%%%%%%%%
%%%%%%%%%%%%%%%%%%%%%%%%%%%%%%%%%%%%%%%%%%%%%%%

%%%%%%%%%%%%%%%%%%%%%%%%%%%%%%%%%%%%%%

\printbibliography

@article{DelPiano:2024nrl,
    author = "Del Piano, Manuel and Hohenegger, Stefan and Sannino, Francesco",
    title = "{Black hole shadow and other observables away from the horizon: Extending the effective metric descriptions}",
    eprint = "2412.13673",
    archivePrefix = "arXiv",
    primaryClass = "gr-qc",
    doi = "10.1103/PhysRevD.111.064070",
    journal = "Phys. Rev. D",
    volume = "111",
    number = "6",
    pages = "064070",
    year = "2025"
}

@Inbook{Rezzolla2016,
author="Rezzolla, Luciano",
editor="Haardt, Francesco
and Gorini, Vittorio
and Moschella, Ugo
and Treves, Aldo
and Colpi, Monica",
title="An Introduction to Astrophysical Black Holes and Their Dynamical Production",
bookTitle="Astrophysical Black Holes",
year="2016",
publisher="Springer International Publishing",
address="Cham",
pages="1--44",
isbn="978-3-319-19416-5",
doi="10.1007/978-3-319-19416-5_1"
}

@article{Jha:2024ltc,
    author = "Jha, Sohan Kumar",
    title = "{Shadow, ISCO, quasinormal modes, Hawking spectrum, weak gravitational lensing, and parameter estimation of a Schwarzschild black hole surrounded by a Dehnen type dark matter halo}",
    eprint = "2407.18509",
    archivePrefix = "arXiv",
    primaryClass = "gr-qc",
    doi = "10.1088/1475-7516/2025/03/054",
    journal = "JCAP",
    volume = "03",
    pages = "054",
    year = "2025"
}

@article{SH,
    author = "Hohenegger, Stefan",
    title = "{Probing effective black hole deformations}",
    eprint = "2508.17781",
    archivePrefix = "arXiv",
    primaryClass = "gr-qc",
    doi = "10.1140/epjc/s10052-025-15126-5",
    journal = "Eur. Phys. J. C",
    volume = "85",
    number = "12",
    pages = "1413",
    year = "2025"
}

@article{DelPiano:2023fiw,
    author = "Del Piano, Manuel and Hohenegger, Stefan and Sannino, Francesco",
    title = "{Quantum black hole physics from the event horizon}",
    eprint = "2307.13489",
    archivePrefix = "arXiv",
    primaryClass = "gr-qc",
    doi = "10.1103/PhysRevD.109.024045",
    journal = "Phys. Rev. D",
    volume = "109",
    number = "2",
    pages = "024045",
    year = "2024"
}

@article{DelPiano:2024gvw,
    author = "Del Piano, Manuel and Hohenegger, Stefan and Sannino, Francesco",
    title = "{Effective metric descriptions of quantum black holes}",
    eprint = "2403.12679",
    archivePrefix = "arXiv",
    primaryClass = "gr-qc",
    doi = "10.1140/epjc/s10052-024-13609-5",
    journal = "Eur. Phys. J. C",
    volume = "84",
    number = "12",
    pages = "1273",
    year = "2024"
}

@book{Hawking:1973uf,
    author = "Hawking, Stephen W. and Ellis, George F. R.",
    title = "{The Large Scale Structure of Space-Time}",
    doi = "10.1017/9781009253161",
    isbn = "978-1-009-25316-1",
    publisher = "Cambridge University Press",
    series = "Cambridge Monographs on Mathematical Physics",
    month = "2",
    year = "1973"
}

@article{Boonserm:2015aqa,
    author = "Boonserm, Petarpa and Ngampitipan, Tritos and Visser, Matt",
    title = "{Mimicking static anisotropic fluid spheres in general relativity}",
    eprint = "1501.07044",
    archivePrefix = "arXiv",
    primaryClass = "gr-qc",
    doi = "10.1142/S021827181650019X",
    journal = "Int. J. Mod. Phys. D",
    volume = "25",
    number = "02",
    pages = "1650019",
    year = "2015"
}

@article{Simpson:2018tsi,
    author = "Simpson, Alex and Visser, Matt",
    title = "{Black-bounce to traversable wormhole}",
    eprint = "1812.07114",
    archivePrefix = "arXiv",
    primaryClass = "gr-qc",
    doi = "10.1088/1475-7516/2019/02/042",
    journal = "JCAP",
    volume = "02",
    pages = "042",
    year = "2019"
}

@article{Simpson:2019mud,
    author = "Simpson, Alex and Visser, Matt",
    title = "{Regular black holes with asymptotically Minkowski cores}",
    eprint = "1911.01020",
    archivePrefix = "arXiv",
    primaryClass = "gr-qc",
    doi = "10.3390/universe6010008",
    journal = "Universe",
    volume = "6",
    number = "1",
    pages = "8",
    year = "2019"
}

@book{Carroll:2004st,
    author = "Carroll, Sean M.",
    title = "{Spacetime and Geometry}: {An Introduction to General Relativity}",
    doi = "10.1017/9781108770385",
    isbn = "978-0-8053-8732-2, 978-1-108-48839-6, 978-1-108-77555-7",
    publisher = "Cambridge University Press",
    month = "7",
    year = "2019"
}

@article{Perlick:2021aok,
    author = "Perlick, Volker and Tsupko, Oleg Yu.",
    title = "{Calculating black hole shadows: Review of analytical studies}",
    eprint = "2105.07101",
    archivePrefix = "arXiv",
    primaryClass = "gr-qc",
    doi = "10.1016/j.physrep.2021.10.004",
    journal = "Phys. Rept.",
    volume = "947",
    pages = "1--39",
    year = "2022"
}

@article{Cardoso:2008bp,
    author = "Cardoso, Vitor and Miranda, Alex S. and Berti, Emanuele and Witek, Helvi and Zanchin, Vilson T.",
    title = "{Geodesic stability, Lyapunov exponents and quasinormal modes}",
    eprint = "0812.1806",
    archivePrefix = "arXiv",
    primaryClass = "hep-th",
    doi = "10.1103/PhysRevD.79.064016",
    journal = "Phys. Rev. D",
    volume = "79",
    number = "6",
    pages = "064016",
    year = "2009"
}

@article{Pretorius:2007jn,
    author = "Pretorius, Frans and Khurana, Deepak",
    editor = "Campanelli, Manuela and Rezzolla, Luciano",
    title = "{Black hole mergers and unstable circular orbits}",
    eprint = "gr-qc/0702084",
    archivePrefix = "arXiv",
    doi = "10.1088/0264-9381/24/12/S07",
    journal = "Class. Quant. Grav.",
    volume = "24",
    pages = "S83--S108",
    year = "2007"
}

@article{Schutz:1985km,
    author = "Schutz, Bernard F. and Will, Clifford M.",
    title = "{Black Hole Normal Modes: a Semianalytic Approach}",
    reportNumber = "PRINT-85-0063 (WASH.U.,ST.LOUIS)",
    doi = "10.1086/184453",
    journal = "Astrophys. J. Lett.",
    volume = "291",
    pages = "L33--L36",
    year = "1985"
}

@article{Iyer:1986np,
    author = "Iyer, Sai and Will, Clifford M.",
    title = "{Black Hole Normal Modes: A {WKB} Approach. 1. Foundations and Application of a Higher Order {WKB} Analysis of Potential Barrier Scattering}",
    reportNumber = "Print-86-1482 (WASH. U., ST. LOUIS)",
    doi = "10.1103/PhysRevD.35.3621",
    journal = "Phys. Rev. D",
    volume = "35",
    pages = "3621",
    year = "1987"
}

@article{Iyer:1986nq,
    author = "Iyer, Sai",
    title = "{Black Hole Normal Modes: a WKB approach. 2. Schwarzschild Black Holes }",
    reportNumber = "Print-86-1483 (WASH. U., ST. LOUIS)",
    doi = "10.1103/PhysRevD.35.3632",
    journal = "Phys. Rev. D",
    volume = "35",
    pages = "3632",
    year = "1987"
}

@article{DelPiano:2025ykr,
    author = "Del Piano, Manuel and De Simone, Ciro and Damia Paciarini, Mattia and De Falco, Vittorio and Myszkowski, Miko{\l}aj and Sannino, Francesco and Vellucci, Vania",
    title = "{Toward a unified view of agnostic parametrizations for deformed black holes}",
    eprint = "2510.14707",
    archivePrefix = "arXiv",
    primaryClass = "gr-qc",
    doi = "10.1103/x9fw-lrv7",
    journal = "Phys. Rev. D",
    volume = "113",
    number = "6",
    pages = "064010",
    year = "2026"
}

@article{Cadoni:2022vsn,
    author = "Cadoni, M. and De Laurentis, M. and De Martino, I. and Della Monica, R. and Oi, M. and Sanna, A. P.",
    title = "{Are nonsingular black holes with super-Planckian hair ruled out by S2 star data?}",
    eprint = "2211.11585",
    archivePrefix = "arXiv",
    primaryClass = "gr-qc",
    doi = "10.1103/PhysRevD.107.044038",
    journal = "Phys. Rev. D",
    volume = "107",
    number = "4",
    pages = "044038",
    year = "2023"
}

@article{Abbott:2016blz,
    author = "Abbott, B. P. and others",
    collaboration = "LIGO Scientific and Virgo",
    title = "{Observation of Gravitational Waves from a Binary Black Hole Merger}",
    journal = "Phys. Rev. Lett.",
    volume = "116",
    pages = "061102",
    year = "2016",
    eprint = "1602.03837",
    archivePrefix = "arXiv",
    primaryClass = "gr-qc",
    doi = "10.1103/PhysRevLett.116.061102"
}

@article{Abbott:2016nmj,
  author = "Abbott, B. P. and others",
  collaboration = "LIGO Scientific and Virgo",
  title = "{Tests of general relativity with GW150914}",
  journal = "Phys. Rev. Lett.",
  volume = "116",
  number = "22",
  pages = "221101",
  year = "2016",
  eprint = "1602.03841",
  archivePrefix = "arXiv",
  primaryClass = "gr-qc",
  doi = "10.1103/PhysRevLett.116.221101"
}

@article{Abbott:2019niy,
    author = "Abbott, B. P. and others",
    collaboration = "LIGO Scientific and Virgo",
    title = "{Tests of general relativity with the binary black hole signals from the LIGO-Virgo catalog GWTC-1}",
    journal = "Phys. Rev. D",
    volume = "100",
    number = "10",
    pages = "104036",
    year = "2019",
    eprint = "1903.04467",
    archivePrefix = "arXiv",
    primaryClass = "gr-qc",
    doi = "10.1103/PhysRevD.100.104036"
}

@article{LIGOScientific:2020tif,
    author = "Abbott, R. and others",
    collaboration = "LIGO Scientific, Virgo",
    title = "{Tests of general relativity with binary black holes from the second LIGO-Virgo gravitational-wave transient catalog}",
    eprint = "2010.14529",
    archivePrefix = "arXiv",
    primaryClass = "gr-qc",
    reportNumber = "LIGO-P2000091",
    doi = "10.1103/PhysRevD.103.122002",
    journal = "Phys. Rev. D",
    volume = "103",
    number = "12",
    pages = "122002",
    year = "2021"
}

@article{EventHorizonTelescope:2019dse,
    author = "Akiyama, Kazunori and others",
    collaboration = "Event Horizon Telescope",
    title = "{First M87 Event Horizon Telescope Results. I. The Shadow of the Supermassive Black Hole}",
    eprint = "1906.11238",
    archivePrefix = "arXiv",
    primaryClass = "astro-ph.GA",
    doi = "10.3847/2041-8213/ab0ec7",
    journal = "Astrophys. J. Lett.",
    volume = "875",
    pages = "L1",
    year = "2019"
}

@article{EventHorizonTelescope:2022wkp,
    author = "Akiyama, Kazunori and others",
    collaboration = "Event Horizon Telescope",
    title = "{First Sagittarius A* Event Horizon Telescope Results. I. The Shadow of the Supermassive Black Hole in the Center of the Milky Way}",
    eprint = "2311.08680",
    archivePrefix = "arXiv",
    primaryClass = "astro-ph.HE",
    doi = "10.3847/2041-8213/ac6674",
    journal = "Astrophys. J. Lett.",
    volume = "930",
    number = "2",
    pages = "L12",
    year = "2022"
}

@article{EventHorizonTelescope:2022xqj,
    author = "Akiyama, Kazunori and others",
    collaboration = "Event Horizon Telescope",
    title = "{First Sagittarius A* Event Horizon Telescope Results. VI. Testing the Black Hole Metric}",
    eprint = "2311.09484",
    archivePrefix = "arXiv",
    primaryClass = "astro-ph.HE",
    reportNumber = "FERMILAB-PUB-22-422-PPD",
    doi = "10.3847/2041-8213/ac6756",
    journal = "Astrophys. J. Lett.",
    volume = "930",
    number = "2",
    pages = "L17",
    year = "2022"
}

@article{Dymnikova:1992ux,
    author = "Dymnikova, I.",
    title = "{Vacuum nonsingular black hole}",
    doi = "10.1007/BF00760226",
    journal = "Gen. Rel. Grav.",
    volume = "24",
    pages = "235--242",
    year = "1992"
}

@article{Hayward:2005gi,
    author = "Hayward, Sean A.",
    title = "{Formation and evaporation of regular black holes}",
    eprint = "gr-qc/0506126",
    archivePrefix = "arXiv",
    doi = "10.1103/PhysRevLett.96.031103",
    journal = "Phys. Rev. Lett.",
    volume = "96",
    pages = "031103",
    year = "2006"
}

@article{Dymnikova:2004qg,
    author = "Dymnikova, Irina and Galaktionov, Evgeny",
    title = "{Stability of a vacuum nonsingular black hole}",
    eprint = "gr-qc/0409049",
    archivePrefix = "arXiv",
    doi = "10.1088/0264-9381/22/12/003",
    journal = "Class. Quant. Grav.",
    volume = "22",
    pages = "2331--2358",
    year = "2005"
}

@article{Garcia95,
  title = {Class of Stationary Axisymmetric Solutions of the Einstein-Maxwell-Dilaton-Axion Field Equations},
  author = {Garc\'{\i}a, Alberto and Galtsov, Dmitri and Kechkin, Oleg},
  journal = {Phys. Rev. Lett.},
  volume = {74},
  issue = {8},
  pages = {1276--1279},
  numpages = {0},
  year = {1995},
  month = {Feb},
  publisher = {American Physical Society},
  doi = {10.1103/PhysRevLett.74.1276}
}

@article{Kazakov:1993ha,
    author = "Kazakov, D. I. and Solodukhin, S. N.",
    title = "{On Quantum deformation of the Schwarzschild solution}",
    eprint = "hep-th/9310150",
    archivePrefix = "arXiv",
    reportNumber = "JINR-E2-93-371",
    doi = "10.1016/S0550-3213(94)80045-6",
    journal = "Nucl. Phys. B",
    volume = "429",
    pages = "153--176",
    year = "1994"
}

@article{Schwarzschild:1916uq,
    author = "Schwarzschild, Karl",
    title = "{On the gravitational field of a mass point according to Einstein's theory}",
    eprint = "physics/9905030",
    archivePrefix = "arXiv",
    journal = "Sitzungsber. Preuss. Akad. Wiss. Berlin (Math. Phys. )",
    volume = "1916",
    pages = "189--196",
    year = "1916"
}

@article{Johannsen_2010,
doi = {10.1088/0004-637X/716/1/187},
url = {https://doi.org/10.1088/0004-637X/716/1/187},
year = {2010},
month = {may},
publisher = {The American Astronomical Society},
volume = {716},
number = {1},
pages = {187},
author = {Johannsen, Tim and Psaltis, Dimitrios},
title = {Testing the No-Hair Theorem with Observations in the Electromagnetic Spectrum. I. Properties of a Quasi-Kerr Spacetime},
journal = {The Astrophysical Journal}
}

@article{PhysRevD.83.124015,
  title = {Metric for rapidly spinning black holes suitable for strong-field tests of the no-hair theorem},
  author = {Johannsen, Tim and Psaltis, Dimitrios},
  journal = {Phys. Rev. D},
  volume = {83},
  issue = {12},
  pages = {124015},
  numpages = {10},
  year = {2011},
  month = {Jun},
  publisher = {American Physical Society},
  doi = {10.1103/PhysRevD.83.124015},
  url = {https://link.aps.org/doi/10.1103/PhysRevD.83.124015}
}

@article{642b3682a1c8a315fd231707, title={Mapping spacetimes with LISA: Inspiral of a test body in a “quasi-Kerr” field}, volume={23}, ISSN={0264-9381}, DOI={10.1088/0264-9381/23/12/013}, number={12}, journal={Classical and Quantum Gravity}, author={Glampedakis, Kostas and Babak, Stanislav}, year={2006}, pages={4167-4188} }

@article{Cardoso:2014rha,
    author = "Cardoso, Vitor and Pani, Paolo and Rico, Joao",
    title = "{On generic parametrizations of spinning black-hole geometries}",
    eprint = "1401.0528",
    archivePrefix = "arXiv",
    primaryClass = "gr-qc",
    doi = "10.1103/PhysRevD.89.064007",
    journal = "Phys. Rev. D",
    volume = "89",
    pages = "064007",
    year = "2014"
}

@article{Rezzolla:2014mua,
    author = "Rezzolla, Luciano and Zhidenko, Alexander",
    title = "{New parametrization for spherically symmetric black holes in metric theories of gravity}",
    eprint = "1407.3086",
    archivePrefix = "arXiv",
    primaryClass = "gr-qc",
    doi = "10.1103/PhysRevD.90.084009",
    journal = "Phys. Rev. D",
    volume = "90",
    number = "8",
    pages = "084009",
    year = "2014"
}

@article{Konoplya:2016jvv,
    author = "Konoplya, Roman and Rezzolla, Luciano and Zhidenko, Alexander",
    title = "{General parametrization of axisymmetric black holes in metric theories of gravity}",
    eprint = "1602.02378",
    archivePrefix = "arXiv",
    primaryClass = "gr-qc",
    doi = "10.1103/PhysRevD.93.064015",
    journal = "Phys. Rev. D",
    volume = "93",
    number = "6",
    pages = "064015",
    year = "2016"
}

@article{Berti:2015itd,
    author = "Berti, Emanuele and others",
    title = "{Testing General Relativity with Present and Future Astrophysical Observations}",
    eprint = "1501.07274",
    archivePrefix = "arXiv",
    primaryClass = "gr-qc",
    doi = "10.1088/0264-9381/32/24/243001",
    journal = "Class. Quant. Grav.",
    volume = "32",
    pages = "243001",
    year = "2015"
}

@article{Binetti:2022xdi,
    author = "Binetti, Emanuele and Del Piano, Manuel and Hohenegger, Stefan and Pezzella, Franco and Sannino, Francesco",
    title = "{Effective theory of quantum black holes}",
    eprint = "2203.13515",
    archivePrefix = "arXiv",
    primaryClass = "gr-qc",
    reportNumber = "CERN-TH-2022-048",
    doi = "10.1103/PhysRevD.106.046006",
    journal = "Phys. Rev. D",
    volume = "106",
    number = "4",
    pages = "046006",
    year = "2022"
}

@article{DAlise:2023hls,
    author = "D'Alise, Alessandra and Fabiano, Giuseppe and Frattulillo, Domenico and Hohenegger, Stefan and Iacobacci, Davide and Pezzella, Franco and Sannino, Francesco",
    title = "{Positivity conditions for generalized Schwarzschild space-times}",
    eprint = "2305.12965",
    archivePrefix = "arXiv",
    primaryClass = "gr-qc",
    reportNumber = "LYCEN 2023-01",
    doi = "10.1103/PhysRevD.108.084042",
    journal = "Phys. Rev. D",
    volume = "108",
    number = "8",
    pages = "084042",
    year = "2023"
}

@article{DamiaPaciarini:2024ssf,
    author = "Damia Paciarini, Mattia and Hohenegger, Stefan and Myszkowski, Mikolaj and Sannino, Francesco",
    title = "{Effective metric description of 2+1 dimensional quantum black holes}",
    eprint = "2412.15960",
    archivePrefix = "arXiv",
    primaryClass = "gr-qc",
    doi = "10.1140/epjc/s10052-025-14286-8",
    journal = "Eur. Phys. J. C",
    volume = "85",
    number = "6",
    pages = "611",
    year = "2025"
}

@article{DamiaPaciarini:2025xjc,
    author = "Damia Paciarini, Mattia and Del Piano, Manuel and Hohenegger, Stefan and Sannino, Francesco",
    title = "{Effective metric description of charged black holes}",
    eprint = "2504.20810",
    archivePrefix = "arXiv",
    primaryClass = "gr-qc",
    doi = "10.1140/epjc/s10052-025-14551-w",
    journal = "Eur. Phys. J. C",
    volume = "85",
    number = "8",
    pages = "848",
    year = "2025"
}

@article{Nollert:1999ji,
    author = "Nollert, Hans-Peter",
    title = "{TOPICAL REVIEW: Quasinormal modes: the characteristic `sound' of black holes and neutron stars}",
    doi = "10.1088/0264-9381/16/12/201",
    journal = "Class. Quant. Grav.",
    volume = "16",
    pages = "R159--R216",
    year = "1999"
}

@article{Regge:1957td,
    author = "Regge, Tullio and Wheeler, John A.",
    title = "{Stability of a Schwarzschild singularity}",
    doi = "10.1103/PhysRev.108.1063",
    journal = "Phys. Rev.",
    volume = "108",
    pages = "1063--1069",
    year = "1957"
}

@article{Zerilli:1970wzz,
    author = "Zerilli, F. J.",
    title = "{Gravitational field of a particle falling in a schwarzschild geometry analyzed in tensor harmonics}",
    doi = "10.1103/PhysRevD.2.2141",
    journal = "Phys. Rev. D",
    volume = "2",
    pages = "2141--2160",
    year = "1970"
}

@article{Zerilli:1970se,
    author = "Zerilli, Frank J.",
    title = "{Effective potential for even parity Regge-Wheeler gravitational perturbation equations}",
    doi = "10.1103/PhysRevLett.24.737",
    journal = "Phys. Rev. Lett.",
    volume = "24",
    pages = "737--738",
    year = "1970"
}

@article{Livine:2024bvo,
    author = "Livine, Etera R. and Montagnon, Clara and Oshita, Naritaka and Roussille, Hugo",
    title = "{Scalar Quasi-Normal Modes of a loop quantum black hole}",
    eprint = "2405.12671",
    archivePrefix = "arXiv",
    primaryClass = "gr-qc",
    reportNumber = "YITP-24-58, RIKEN-iTHEMS-Report-24",
    doi = "10.1088/1475-7516/2024/10/037",
    journal = "JCAP",
    volume = "10",
    pages = "037",
    year = "2024"
}

@article{Kokkotas:1999bd,
    author = "Kokkotas, Kostas D. and Schmidt, Bernd G.",
    title = "{Quasinormal modes of stars and black holes}",
    eprint = "gr-qc/9909058",
    archivePrefix = "arXiv",
    doi = "10.12942/lrr-1999-2",
    journal = "Living Rev. Rel.",
    volume = "2",
    pages = "2",
    year = "1999"
}

@article{Berti:2009kk,
    author = "Berti, Emanuele and Cardoso, Vitor and Starinets, Andrei O.",
    title = "{Quasinormal modes of black holes and black branes}",
    eprint = "0905.2975",
    archivePrefix = "arXiv",
    primaryClass = "gr-qc",
    doi = "10.1088/0264-9381/26/16/163001",
    journal = "Class. Quant. Grav.",
    volume = "26",
    pages = "163001",
    year = "2009"
}

@article{Konoplya:2011qq,
    author = "Konoplya, R. A. and Zhidenko, A.",
    title = "{Quasinormal modes of black holes: From astrophysics to string theory}",
    eprint = "1102.4014",
    archivePrefix = "arXiv",
    primaryClass = "gr-qc",
    doi = "10.1103/RevModPhys.83.793",
    journal = "Rev. Mod. Phys.",
    volume = "83",
    pages = "793--836",
    year = "2011"
}

@article{Leaver:1985ax,
    author = "Leaver, E. W.",
    title = "{An Analytic representation for the quasi normal modes of Kerr black holes}",
    doi = "10.1098/rspa.1985.0119",
    journal = "Proc. Roy. Soc. Lond. A",
    volume = "402",
    pages = "285--298",
    year = "1985"
}

@article{Cho:2009cj,
    author = "Cho, H. T. and Cornell, A. S. and Doukas, Jason and Naylor, Wade",
    title = "{Black hole quasinormal modes using the asymptotic iteration method}",
    eprint = "0912.2740",
    archivePrefix = "arXiv",
    primaryClass = "gr-qc",
    reportNumber = "YITP-09-110, WITS-CTP-049",
    doi = "10.1088/0264-9381/27/15/155004",
    journal = "Class. Quant. Grav.",
    volume = "27",
    pages = "155004",
    year = "2010"
}

@article{Cho:2011sf,
    author = "Cho, H. T. and Cornell, A. S. and Doukas, Jason and Huang, T. R. and Naylor, Wade",
    title = "{A New Approach to Black Hole Quasinormal Modes: A Review of the Asymptotic Iteration Method}",
    eprint = "1111.5024",
    archivePrefix = "arXiv",
    primaryClass = "gr-qc",
    reportNumber = "YITP-11-97, WITS-CTP-83, OU-HET-735-2011",
    doi = "10.1155/2012/281705",
    journal = "Adv. Math. Phys.",
    volume = "2012",
    pages = "281705",
    year = "2012"
}

@article{Chen:2017kqa,
    author = "Chen, C. -H. and Cho, H. T. and Cornell, A. S. and Harmsen, G. and Ngcobo, X.",
    title = {{Quasinormal modes and absorption probabilities of spin- 3/2 fields in D -dimensional Reissner-Nordstr{\"o}m black hole spacetimes}},
    eprint = "1710.08024",
    archivePrefix = "arXiv",
    primaryClass = "gr-qc",
    doi = "10.1103/PhysRevD.97.024038",
    journal = "Phys. Rev. D",
    volume = "97",
    number = "2",
    pages = "024038",
    year = "2018"
}

@article{Konoplya:2003ii,
    author = "Konoplya, R. A.",
    title = "{Quasinormal behavior of the d-dimensional Schwarzschild black hole and higher order WKB approach}",
    eprint = "gr-qc/0303052",
    archivePrefix = "arXiv",
    doi = "10.1103/PhysRevD.68.024018",
    journal = "Phys. Rev. D",
    volume = "68",
    pages = "024018",
    year = "2003"
}

@book{Misner:1973prb,
    author = "Misner, Charles W. and Thorne, K. S. and Wheeler, J. A.",
    title = "{Gravitation}",
    isbn = "978-0-7167-0344-0, 978-0-691-17779-3",
    publisher = "W. H. Freeman",
    address = "San Francisco",
    year = "1973"
}

@article{Kubo,
author = {Kubo ,Ryogo},
title = {Statistical-Mechanical Theory of Irreversible Processes. I. General Theory and Simple Applications to Magnetic and Conduction Problems},
journal = {Journal of the Physical Society of Japan},
volume = {12},
number = {6},
pages = {570-586},
year = {1957},
doi = {10.1143/JPSJ.12.570},

eprint = { 
    
        https://doi.org/10.1143/JPSJ.12.570
    
    

}

}

@article{Green,
    author = {Green, Melville S.},
    title = {Markoff Random Processes and the Statistical Mechanics of Time‐Dependent Phenomena. II. Irreversible Processes in Fluids},
    journal = {The Journal of Chemical Physics},
    volume = {22},
    number = {3},
    pages = {398-413},
    year = {1954},
    month = {03},
    issn = {0021-9606},
    doi = {10.1063/1.1740082},
    url = {https://doi.org/10.1063/1.1740082},
    eprint = {https://pubs.aip.org/aip/jcp/article-pdf/22/3/398/18804401/398_1_online.pdf},
}

@article{Streda_1982,
doi = {10.1088/0022-3719/15/22/005},
url = {https://doi.org/10.1088/0022-3719/15/22/005},
year = {1982},
month = {aug},
publisher = {},
volume = {15},
number = {22},
pages = {L717},
author = {P Streda},
title = {Theory of quantised Hall conductivity in two dimensions},
journal = {Journal of Physics C: Solid State Physics}
}

@article{Maldacena:1997re,
    author = "Maldacena, Juan Martin",
    title = "{The Large $N$ limit of superconformal field theories and supergravity}",
    eprint = "hep-th/9711200",
    archivePrefix = "arXiv",
    reportNumber = "HUTP-97-A097, HUTP-98-A097",
    doi = "10.4310/ATMP.1998.v2.n2.a1",
    journal = "Adv. Theor. Math. Phys.",
    volume = "2",
    pages = "231--252",
    year = "1998"
}

@article{Matyjasek:2026yiu,
    author = "Matyjasek, Jerzy and Konoplya, Roman A. and Zhidenko, Alexander",
    title = "{An Efficient Higher-Order WKB Code for Quasinormal Modes and Greybody Factors}",
    eprint = "2603.12466",
    archivePrefix = "arXiv",
    primaryClass = "gr-qc",
    doi = "10.53941/ijgtp.2026.100005",
    journal = "Int. J. Grav. Theor. Phys.",
    volume = "2",
    number = "1",
    pages = "5",
    year = "2026"
}

@article{Miyachi:2025ptm,
    author = "Miyachi, Taiga and Namba, Ryo and Omiya, Hidetoshi and Oshita, Naritaka",
    title = "{Path to an exact WKB analysis of black hole quasinormal modes}",
    eprint = "2503.17245",
    archivePrefix = "arXiv",
    primaryClass = "hep-th",
    reportNumber = "RIKEN-iTHEMS-Report-25",
    doi = "10.1103/1gmr-9f1g",
    journal = "Phys. Rev. D",
    volume = "111",
    number = "12",
    pages = "124045",
    year = "2025"
}

@article{Matyjasek:2017psv,
    author = "Matyjasek, Jerzy and Opala, Micha{\l}",
    title = "{Quasinormal modes of black holes. The improved semianalytic approach}",
    eprint = "1704.00361",
    archivePrefix = "arXiv",
    primaryClass = "gr-qc",
    doi = "10.1103/PhysRevD.96.024011",
    journal = "Phys. Rev. D",
    volume = "96",
    number = "2",
    pages = "024011",
    year = "2017"
}

@article{Konoplya:2019hlu,
    author = "Konoplya, R. A. and Zhidenko, A. and Zinhailo, A. F.",
    title = "{Higher order WKB formula for quasinormal modes and grey-body factors: recipes for quick and accurate calculations}",
    eprint = "1904.10333",
    archivePrefix = "arXiv",
    primaryClass = "gr-qc",
    doi = "10.1088/1361-6382/ab2e25",
    journal = "Class. Quant. Grav.",
    volume = "36",
    pages = "155002",
    year = "2019"
}

@article{Cardoso:2019mqo,
    author = "Cardoso, Vitor and Kimura, Masashi and Maselli, Andrea and Berti, Emanuele and Macedo, Caio F. B. and McManus, Ryan",
    title = "{Parametrized black hole quasinormal ringdown: Decoupled equations for nonrotating black holes}",
    eprint = "1901.01265",
    archivePrefix = "arXiv",
    primaryClass = "gr-qc",
    doi = "10.1103/PhysRevD.99.104077",
    journal = "Phys. Rev. D",
    volume = "99",
    number = "10",
    pages = "104077",
    year = "2019"
}

@article{McManus:2019ulj,
    author = "McManus, Ryan and Berti, Emanuele and Macedo, Caio F. B. and Kimura, Masashi and Maselli, Andrea and Cardoso, Vitor",
    title = "{Parametrized black hole quasinormal ringdown. II. Coupled equations and quadratic corrections for nonrotating black holes}",
    eprint = "1906.05155",
    archivePrefix = "arXiv",
    primaryClass = "gr-qc",
    doi = "10.1103/PhysRevD.100.044061",
    journal = "Phys. Rev. D",
    volume = "100",
    number = "4",
    pages = "044061",
    year = "2019"
}

@article{EventHorizonTelescope:2022urf,
    author = "Akiyama, Kazunori and others",
    collaboration = "Event Horizon Telescope",
    title = "{First Sagittarius A* Event Horizon Telescope Results. V. Testing Astrophysical Models of the Galactic Center Black Hole}",
    eprint = "2311.09478",
    archivePrefix = "arXiv",
    primaryClass = "astro-ph.HE",
    reportNumber = "FERMILAB-PUB-22-419-PPD",
    doi = "10.3847/2041-8213/ac6672",
    journal = "Astrophys. J. Lett.",
    volume = "930",
    number = "2",
    pages = "L16",
    year = "2022"
}

@article{EventHorizonTelescope:2021srq,
    author = "Akiyama, Kazunori and others",
    collaboration = "Event Horizon Telescope",
    title = "{First M87 Event Horizon Telescope Results. VIII. Magnetic Field Structure near The Event Horizon}",
    eprint = "2105.01173",
    archivePrefix = "arXiv",
    primaryClass = "astro-ph.HE",
    reportNumber = "FERMILAB-PUB-21-850-PPD",
    doi = "10.3847/2041-8213/abe4de",
    journal = "Astrophys. J. Lett.",
    volume = "910",
    number = "1",
    pages = "L13",
    year = "2021"
}

@article{EventHorizonTelescope:2021bee,
    author = "Akiyama, Kazunori and others",
    collaboration = "Event Horizon Telescope",
    title = "{First M87 Event Horizon Telescope Results. VII. Polarization of the Ring}",
    eprint = "2105.01169",
    archivePrefix = "arXiv",
    primaryClass = "astro-ph.HE",
    reportNumber = "FERMILAB-PUB-21-849-PPD",
    doi = "10.3847/2041-8213/abe71d",
    journal = "Astrophys. J. Lett.",
    volume = "910",
    number = "1",
    pages = "L12",
    year = "2021"
}

@article{Myszkowski:2025hdc,
    author = "Myszkowski, Miko{\l}aj and Damia Paciarini, Mattia and Sannino, Francesco and Vellucci, Vania",
    title = "{Regular spacetimes in the effective metric description}",
    eprint = "2506.12620",
    archivePrefix = "arXiv",
    primaryClass = "gr-qc",
    doi = "10.1140/epjc/s10052-026-15525-2",
    journal = "Eur. Phys. J. C",
    volume = "86",
    number = "4",
    pages = "346",
    year = "2026"
}

\end{document}